\documentclass[a4paper,11pt]{article}
\pdfoutput=1
\usepackage{jheppub}

\usepackage{amsmath,latexsym,amssymb,slashed}
\usepackage{graphicx}
\usepackage{psfrag}
\usepackage[latin1]{inputenc}
\usepackage{subfig}
\usepackage{slashed}
\usepackage{tensor}
\usepackage{bbm}

\usepackage{float}
\usepackage{placeins}

\usepackage{tabularx}

\usepackage{psfrag}
\usepackage{tikz} 
\usetikzlibrary{arrows}
\usetikzlibrary{plotmarks}
\usetikzlibrary{external}
\usepackage{pgfplots}
\pgfplotsset{compat=1.9}
\usetikzlibrary{patterns}
\usetikzlibrary{calc}
\usepackage{indentfirst}

\newcommand\beal{\begin{align}}

\newcommand{\seff}{\mathcal{S}_{\text{eff}}}

\newcommand{\eq}[1]{\begin{equation}#1\end{equation}}
\newcommand{\spl}[1]{\begin{split}#1\end{split}}

\renewcommand{\t}{\theta}

\newcommand{\boxedeq}[1]{
\begin{equation}
\fbox{
\rule[0.7cm]{0pt}{0pt}
$#1$
\rule[-0.45cm]{0pt}{0pt}
}
\end{equation}
}

\def\slashchar#1{\setbox0=\hbox{$#1$}           
\dimen0=\wd0                                 
\setbox1=\hbox{/} \dimen1=\wd1               
\ifdim\dimen0>\dimen1                        
\rlap{\hbox to \dimen0{\hfil/\hfil}}      
#1                                        
\else                                        
\rlap{\hbox to \dimen1{\hfil$#1$\hfil}}   
/                                         
\fi}

\def\Re           {{\rm Re\hskip0.1em}}

\usepackage{xcolor}
\usepackage{color}

\newcommand{\cnote}[1]{}

\pdfoutput=1 

\usepackage[T1]{fontenc} 

\title{\boldmath One-loop bosonic string and de Sitter space}

\author[a,b]{Sabine Harribey}  
\author[c]{and Dimitrios Tsimpis}

\affiliation[a]{Centre de Physique Th\'{e}orique \\
\'{E}cole Polytechnique, 91128 Palaiseau Cedex, France
}

\affiliation[b]{\'{E}cole Normale Sup\'{e}rieure de Lyon \\
46 All\'{e}e d'Italie, 69007 Lyon, France}

\affiliation[c]{Universit\'{e} de Lyon\\
UCBL, Institut de Physique Nucl\'{e}aire de Lyon\\
UMR 5822, CNRS/IN2P3\\
4 rue Enrico Fermi,
69622 Villeurbanne Cedex,  France\\}

\emailAdd{sabine.harribey@ens-lyon.fr,tsimpis@ipnl.in2p3.fr}

\abstract{We calculate the bosonic string one-loop three- and four-point amplitudes to quadradic order in momentum, and we read off 
 the one-loop low-energy two-derivative effective action for the massless fields, $\seff$.  
Treating the renormalized one-loop vacuum energy as a tunable parameter and extrapolating to a supercritical dimension $D>26$, one can reach a regime where the one-loop 
couplings in $\seff$ are of the same order as the tree-level ones while all higher-loop corrections are negligible. Moreover the effective spacetime curvature is small 
in string units. 
We show that the effective action thus obtained admits weakly-curved de Sitter solutions with constant dilaton at small string coupling.}

\begin{document}
\maketitle
\flushbottom
\setcounter{footnote}{0}
\renewcommand{\thefootnote}{\arabic{footnote}}
\setcounter{section}{0}
\vfill\break

\section{Introduction and summary}

The two-derivative tree-level effective action of bosonic string theory has been extensively used for cosmological applications: it is the starting point of what is 
sometimes called ``tree-level string cosmology'' \cite{Antoniadis:1988aa,Antoniadis:1988vi,Antoniadis:1990uu,Veneziano:1991ek,Tseytlin:1991xk,Tseytlin:1991ss,Tseytlin:1992ye,Tseytlin:1992jq,Gasperini:1992em,Tseytlin:1994cd,Copeland:1994vi,Easther:1995ba}, which exploits the presence of a tree-level cosmological constant in noncritical dimensions. 
In the present paper we go beyond tree level  (in the string coupling) and calculate the bosonic string  one-loop three- and four-point amplitudes  to quadradic order in momentum. 
To our knowledge these results have not appeared before in the literature. 
From that, we are then able to extract the 
one-loop low-energy two-derivative effective action $\seff$ for the massless fields.

By definition, $\seff$ is an action in target space whose tree-level amplitudes reproduce the full one-loop string theory amplitude for the massless fields: the graviton, the antisymmetric two-form  and the dilaton, and can thus be read off systematically from the string theory amplitude.  
While the construction of the effective action is in principle straightforward (albeit technically involved) for the superstring,  
there are two additional complications which arise in the case of  the bosonic string:  the nonvanishing dilaton tadpole (which is generally present in nonsupersymmetric string models), and the tachyon.

As is well known, the presence of the tachyon in the spectrum signals an instability of the vacuum. We do not offer a way to circumvent this problem: as in most of the literature on 
the subject (see however \cite{Tseytlin:1991bu}) we will  
concentrate on the massless fields alone, simply ignoring the tachyonic couplings in the effective action. 
An additional issue with the tachyon is that it gives an infinite contribution to the one-loop vacuum energy, resulting in an infinite cosmological term. The way we 
deal with this here is to simply renormalize by hand the value of the one-loop vacuum energy to a finite value $\Lambda$, which we will treat as a tunable parameter of the bosonic string model.\footnote{A numerical estimate for $\Lambda$, obtained by removing the divergent tachyon contribution in $D=26$, is given in appendix \ref{app:numest}; it depends crucially on the 
 ratio of gravitational (Planck) to string length.} On the other hand, there are numerous  tachyon-free string theory models which are non-supersymmetric and therefore are expected to generally develop 
a cosmological constant at one loop. Our methods are readily transferable to the study of these, potentially more realistic, models.

\begin{figure}[tb!]
\begin{center}
\includegraphics[width=0.65\linewidth]{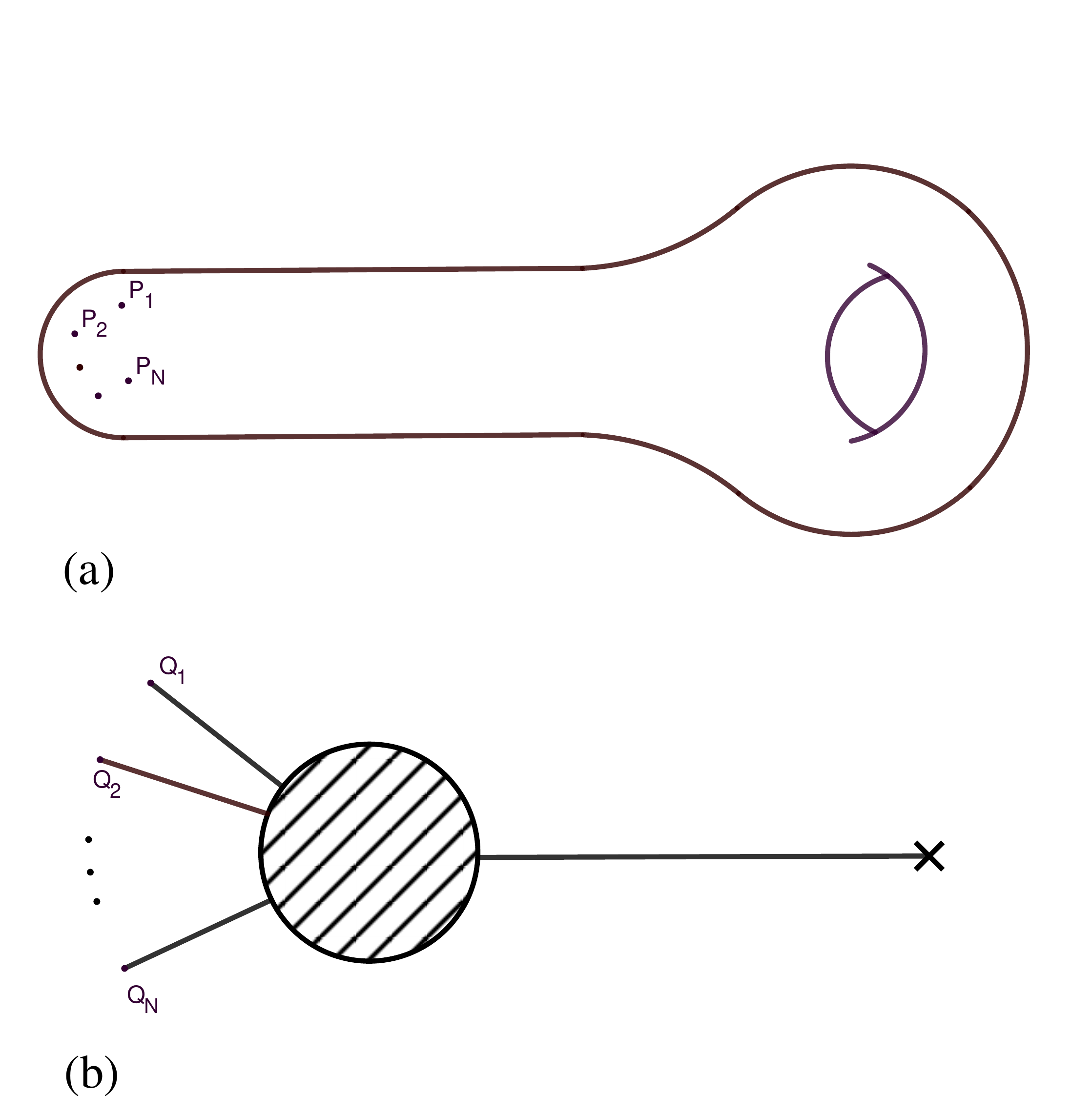}
\end{center}
\caption{(a) String-theory $N$-point amplitude with $N$ vertex operator insertions coming together at the end of a long cylinder. (b)  Its low-energy field-theory limit corresponding to  tree-level diagrams with a propagator going into the tadpole (represented by a cross) at zero momentum.} \label{urfig1}
\end{figure}

\begin{figure}[tb!]
\begin{center}
\includegraphics[width=0.65\linewidth]{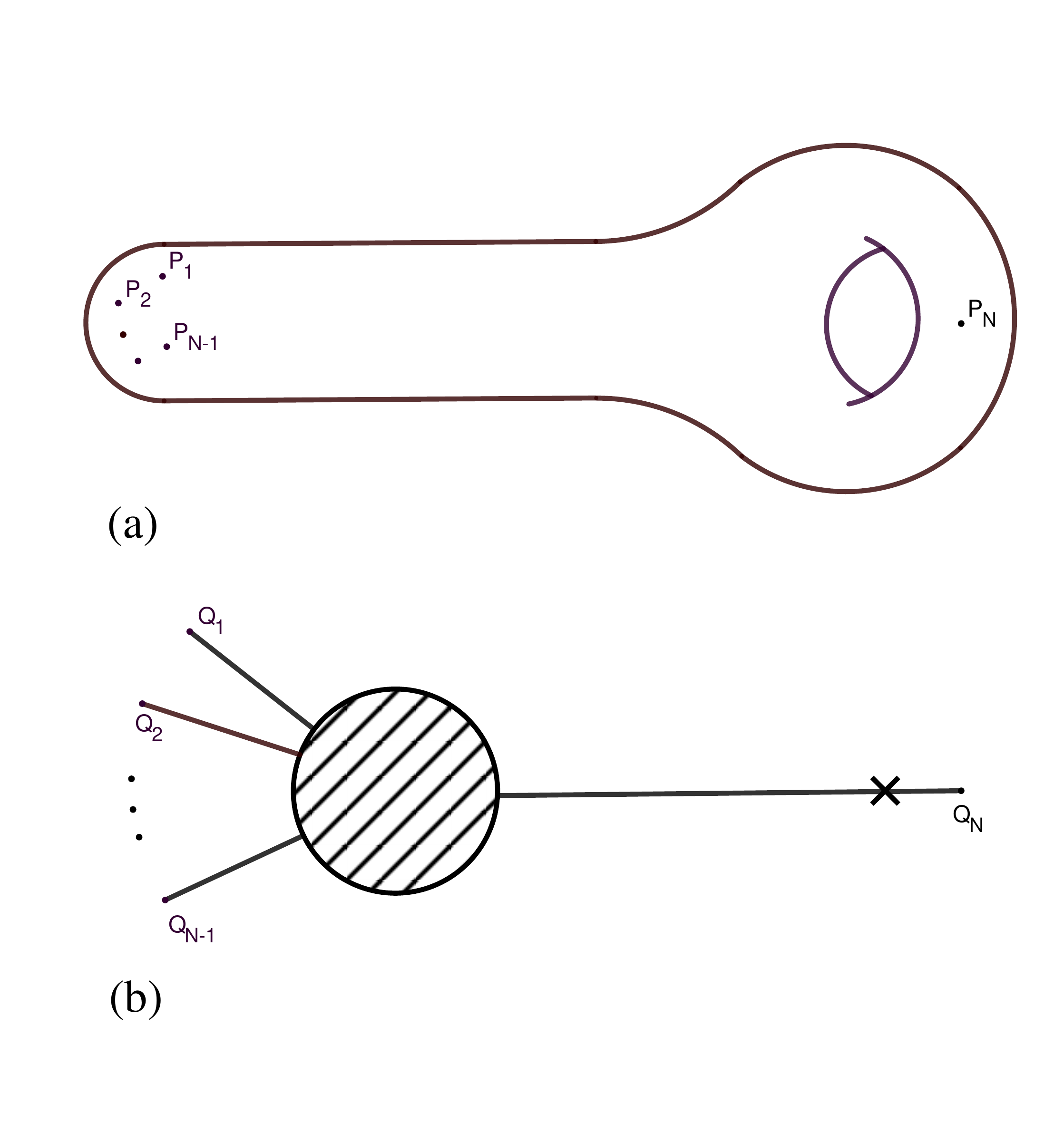}
\end{center}
\caption{String-theory $N$-point amplitude with $N-1$ vertex operator insertions coming together at the end of a long cylinder. (b)  Its low-energy field-theory limit corresponding to  tree-level diagrams with a propagator going on shell due to momentum conservation.} \label{urfig2}
\end{figure}

The dilaton tadpole reflects the fact that one is expanding around the wrong vacuum. 
It is related to the appearance of a non-vanishing cosmological constant  at one-loop order in the string coupling --which is otherwise a desirable feature with regards to  
potential cosmological applications.  
In the presence of tadpoles momentum-independent infinities arise in the two types of  amplitudes depicted in figs.~\ref{urfig1},~\ref{urfig2}. (These are one-loop diagrams, but the argument can be generalized to arbitrary  order in the coupling expansion). For an $N$-point string amplitude, the infinities arise whenever $N$ or $N-1$ vertex operator insertions come together at the end of a long cylinder, as in figs.~\ref{urfig1}a,~\ref{urfig2}a respectively. String perturbation becomes cumbersome in the presence of tadpoles, nevertheless 
$\seff$ is expected to remain a well-defined object \cite{Tseytlin:1988mw}. From the point of view of the low-energy 
effective action $\seff$, these string diagrams correspond to tree-level Feynman diagrams of the type depicted in figs.~\ref{urfig1}b,~\ref{urfig2}b: 1-particle reducible diagrams 
where the propagator of a massless field either goes into the tadpole at zero momentum  (fig.~\ref{urfig1}b), or becomes on-shell due to momentum conservation (fig.~\ref{urfig2}b), leading to $\lim_{\varepsilon\rightarrow0}\tfrac{1}{\varepsilon}$ factors. These are well-understood IR divergences which can be treated within the framework of the low-energy effective action.
As we will see, subtracting the contribution of these 1-particle reducible diagrams from the string amplitude removes the IR divergences and leads to a well-defined two-derivative low-energy effective action,
\boxedeq{\spl{\label{sefffinresc1}
\kappa^2\seff=  
\int\text{d}^Dx\sqrt{G}&\Big\{
(1+\gamma ~\!e^{\sqrt{D-2}~\!\!\phi})\big(
 \tfrac{1}{2}R-\tfrac12 (\partial\phi)^2\big)\\
-\tfrac16 &(1+3\gamma  ~\!e^{\sqrt{D-2}~\!\!\phi}) H^2e^{-\frac{4}{\sqrt{D-2}}~\!\!\phi}
+
\tfrac{1}{\alpha'}e^{\frac{2}{\sqrt{D-2}}~\!\!\phi}(\delta+\alpha'\Lambda e^{\sqrt{D-2}~\!\!\phi})
\Big\}
~,}}
%
%
where $\kappa$ is  the gravitational coupling constant. Moreover 
 we have extrapolated off the critical dimension and included the tree-level cosmological term,
\eq{\label{dlt}\delta:=\frac{26-D}{3}~,}
which vanishes in the critical dimension $D=26$; $\Lambda$ is the renormalized (finite) one-loop vacuum expectation value, cf.~section \ref{sec:ampl}. Besides the cosmological tree-level and one-loop terms, the two-derivative effective action 
(\ref{sefffinresc1}) contains the canonical (Einstein-frame) tree-level kinetic terms for the massless fields plus their one-loop corrections, which are proportional to the dimensionless constant 
$\gamma$ given by, 
\eq{\label{cdef1}
\gamma:=c\alpha' \Lambda
~.}
The constant $c$ can be thought of as the renormalized part of the Eisenstein series $E_1(\tau)$, cf.~(\ref{r2c}) below.  The precise value of $c$ is not important for our purposes (cf.~appendix \ref{app:numest} for a numerical estimate): it is obtained from the one-loop three-point amplitude by subtracting the IR divergences of the type depicted in figs.~\ref{urfig1}, \ref{urfig2}.  
Indeed, as we will see, the three-point amplitude at quadratic momentum is proportional to the Eisenstein series $E_1(\tau)$ (before integration over the modulus of the torus), which has a singularity of the  form $\lim_{\varepsilon\to0}\tfrac{1}{\varepsilon}$. Consequently, upon integration over the fundamental domain of the torus, the singular part of 
 three-point amplitude turns out to be proportional to $\tfrac{1}{\varepsilon}\Lambda$, in agreement with the expected IR divergences of figs.~\ref{urfig1}, \ref{urfig2}.

In our conventions, the closed string coupling constant $g_{\text{str}}$ is related to the vacuum expectation value of the canonically-normalized Einstein-frame dilaton, $\phi$,  via,
\eq{\label{gstr}
g_{\text{str}}^2=\exp\left({\sqrt{D-2}~\!\langle\phi\rangle}\right)
~.}
In general the one-loop corrections considered here would be subject to the usual objection that when quantum corrections can be computed they 
are too small to make any qualitative difference while when they are  important their computation cannot be trusted since perturbation breaks down. 
However, in the  present case, there is a crucial caveat to that 
statement: if the renormalized one-loop vacuum energy can be treated as a tunable parameter,  one could take $\alpha' \Lambda$ to be sufficiently large, 
while at the same time $g_{\text{str}}\ll1$ and perturbation theory remains valid. In this way one can be in a regime where the one-loop corrections  in the effective action are of the same order as the tree-level couplings,
\eq{\label{regime}
\alpha' \Lambda\gg1~;~~~g_{\text{str}}\ll 1~;~~~
\alpha'\Lambda e^{\sqrt{D-2}~\!\!\phi} \sim\mathcal{O}(1)
~.}
Similarly to the one-loop effective action (\ref{sefffinresc1}), the general form of the two-derivative effective action at $k$-th order in string perturbation will 
be given  by couplings of the order of $\alpha'\Lambda_k ~\!e^{k\sqrt{D-2}~\!\phi}$, where $\Lambda_k$ is the $k$-loop vacuum energy. 
It is plausible to assume that, without fine tuning, the  higher-loop vacuum energies will be of the same order as the one-loop ones,
\eq{\label{highrt}\Lambda\sim\Lambda_k,~~~k=2,3,\dots~.}
One is then justified to ignore  higher-order corrections to the effective action (\ref{sefffinresc1}) since these will be suppressed by additional powers of $g_{\text{str}}^2\sim e^{\sqrt{D-2}~\!\!\phi}$. Moreover the cosmological tree-level and one-loop terms in \eqref{sefffinresc1} are  multiplied by an overall $e^{\frac{2}{\sqrt{D-2}}~\!\!\phi}$ factor, which goes like a positive power of $g_{\text{str}}$. Therefore these terms are suppressed in the regime \eqref{regime}, rendering the effective cosmological constant small in units of $\alpha'$.

$\bullet$ {\it de Sitter solutions}

For an infinite range of values of the renormalized constant $c$ in (\ref{cdef1}), the effective action (\ref{sefffinresc1}) admits 
simple $D$-dimensional weakly-curved  de Sitter solutions in a regime where (\ref{regime}) is valid. 
These solutions have vanishing fieldstrength for the antisymmetric two-form, $H=0$, and constant dilaton, $\phi=\text{const}$. 
The dilaton is given by,
\begin{equation}
  e^{\sqrt{D-2}~\!\!\phi}=\frac{2}{\alpha'\Lambda}  
\left( 
c(D-2)+\frac{3D}{(D-26)}
\right)^{-1}~.
\end{equation}
We see that $g^2_{\text{str}}= e^{\sqrt{D-2}~\!\!\phi}$ can be made as small as desired by tuning  $\alpha'\Lambda$ to be sufficiently large. 
Moreover, the curvature, $\lambda$, of  the  de Sitter space is given by,
\begin{equation}
\lambda=\tfrac{2(D-26)}{3\alpha'D}~\!e^{\frac{2}{\sqrt{D-2}}\phi}~,
\end{equation}
which is positive for $D>26$. 
Provided $g^2_{\text{str}}$ is small (which, as we mentioned, can be achieved by tuning  $\alpha'\Lambda$ to be sufficiently large), 
the de Sitter space  is weakly curved, $\lambda\alpha'\ll1$, so that the supergravity solution can be trusted. Let us  stress that even without fine-tuning of $c$, $\Lambda$,  
we can have solutions with small de Sitter curvature and $g^2_{\text{str}}$ in the perturbative regime.

$\bullet$ {\it Outline of the paper}

In section \ref{sec:npoint} we discuss the vanishing-momentum limit of the $N$-point one-loop amplitude. In section \ref{sec:3p2k} we calculate the three-point amplitude 
at quadratic order in momentum, given in \eqref{3s}, \eqref{2b} below. Although the on-shell massless three-point amplitude  vanishes identically for kinematical reasons, it has been known since \cite{Gross:1986mw} that the effective action can still be read off of it using a formal procedure. The effective one-loop, two-derivative action is extracted from the three-point amplitude in section \ref{sec:eff}. 
In section \ref{sec:4p2k} we calculate the four-point amplitude 
at quadratic order in momentum, given in \eqref{a4k2} below. 
Contrary to the three-point amplitude which only exhibits the momentum-independent IR singularities of figs.~\ref{urfig1}, \ref{urfig2}, the  four-point amplitude 
has numerous additional momentum-dependent singularities. 
From the point of view of the low-energy effective action, these 
correspond  to additional 1-particle reducible diagrams that need to be subtracted in order to read off the four-point couplings of the action. 
As in the three-point case,  the four-point amplitude at quadratic momentum turns out to be proportional to the Eisenstein series $E_1(\tau)$. 
Consequently, subtracting the pole singularity  thereof  renormalizes at the same time both the momentum-dependent and the momentum-independent 
singularities of the four-point amplitude (this is different from the regularization of \cite{Green:1999:2} which is obtained by a cutoff of the fundamental domain of the torus). 
We do not attempt a complete comparison of the four-point amplitude with the effective action \eqref{sefffinresc1}, although  
we do show in section \ref{sec:comp} that a certain subset of the terms in the amplitude are consistent with $\seff$. 
In section \ref{sec:desitter} we show that the equations of motion resulting from the effective action \eqref{sefffinresc1} admit simple  weakly-curved de Sitter solutions with constant dilaton at small string coupling. We conclude in section \ref{sec:conclusions}. 
Appendix \ref{app:def} contains our conventions for the special functions used in the main text. A review of the main formul\ae{} used in the calculation of the one-loop amplitude is included in appendix \ref{sec:amplitude}. 
Some prescriptions leading to numerical estimates for the renormalized constants $c$, $\Lambda$ are discussed in 
appendix \ref{app:numest}.



\section{One-loop $N$-point amplitude}\label{sec:npoint}

We refer to appendix \ref{sec:amplitude} for a review of our notations and conventions. 
Following \cite{Kawai:1985xq}, it is convenient to represent the massless vertex operator (\ref{masslessv}) as follows,
\eq{\label{masslessvb}V_i(z_i,\bar{z}_i)=
 \left.
e^{ik_i\cdot X +  \xi_i\cdot\partial X+\bar{\xi}_i\cdot\bar{\partial}X}\right|_{(\xi_i^{\mu},\bar{\xi}_i^{\nu})\rightarrow\xi_i^{\mu\nu}}
~,}
where the notation means that we are to Taylor-expand the exponential, keep  the bilinear term in $\xi_i^{\mu}$, $\bar{\xi}_i^{\nu}$  and make the replacement   
$(\xi_i^{\mu}$, $\bar{\xi}_i^{\nu})\rightarrow\xi_i^{\mu\nu}$. 
(Note that only the polarizations $\xi_i^{\mu\nu}$ are physical: the $(\xi_i^{\mu}$, $\bar{\xi}_i^{\nu})$ are only used in intermediate steps 
as a convenient calculational device.) 
The above then leads to the following formula for correlator of $N$ vertex operators,
\eq{\spl{\label{correlt}
\langle \big[V_1  \big]_R\dots
 \big[V_N  \big]_R\rangle = \prod_{i<j} e^{-k_i\cdot k_j G_{ij}}&\Big\{
 S_{1,\dots, N}\\
 &-\frac{\alpha'}{8\pi\tau_2} (
 \xi_{1}S_{2,\dots, N}+
 \xi_{2}S_{1,3,\dots, N}+\dots+
 \xi_{N}S_{1,\dots, N-1}
 )
 \\
 &+\Big(\frac{\alpha'}{8\pi\tau_2} \Big)^2(
  \xi_{1}\xi_{2}S_{3,\dots, N}+
  \xi_{1}\xi_{3}S_{2,4,\dots, N}+
  \dots+
   \xi_{N-1}\xi_{N}S_{1,\dots, N-2}
 )\\
 &+\dots\\
 &+(-1)^N\Big(\frac{\alpha'}{8\pi\tau_2} \Big)^{N} \xi_{1}\cdots\xi_{N}
\Big\}~,}}
where,
\eq{\label{correlt2}
S_{1,\dots, N} :=
\exp\Big\{
\sum_{i<j}(w_{ij}+w_{\bar{i}\bar{j}}+w_{\bar{i}j }+  w_{i\bar{j} })
+
\sum_i 
\left.
(u_{i }+  u_{\bar{i}})
\Big\}
\right|_{(\xi_i^{\mu},\bar{\xi}_i^{\nu})\rightarrow\xi_i^{\mu\nu}}
~,}
and we have taken (\ref{regv}),(\ref{16}) into account.

\subsection{One- and two-point amplitudes}\label{sec:ampl}

The one-, two- and three-point torus amplitudes at vanishing momentum are simple enough to be able to compute without resorting to the diagrammatic techniques 
explained later in section \ref{sec:32}. Here  we give some more details of the calculation.

Specializing to the case $N=1$, we obtain the one-point dilaton correlator at zero momentum: 
\eq{\label{19o}\langle [V(z_1,\bar{z}_1)]_R\rangle=-
\frac{\alpha'}{8\pi\tau_2}\xi_{1\mu}{}^{\mu}
~,}
where we have taken (\ref{regv}) into account. 
Specializing to the case $N=2$ and taking all contractions, we obtain the two-point graviton, dilaton and antisymmetric tensor correlators:
\eq{\label{twopt}\langle [V(z_1,\bar{z}_1) ]_R [V(z_2,\bar{z}_2)]_R\rangle=
w_{12}w_{\bar{1}\bar{2}}+w_{\bar{1}2}w_{1\bar{2}}+\xi_{1\mu}{}^{\mu}\xi_{2\nu}{}^{\nu}\left(
\frac{\alpha'}{8\pi\tau_2}
\right)^2
~,}
where we have 
taken into account the momentum conservation relation, $k_1+k_2=0$, the on-shell mass condition, $k_i^2=0$, and the transversality of the polarization,  $k_{\mu}\xi^{\mu\nu}=0$, which together imply that all potential momentum-dependent terms in the two-point amplitude vanish. Moreover taking (\ref{15}),(\ref{16}) into account we obtain:
\eq{\spl{\label{19}\frac{1}{\alpha^{\prime2}}\langle [V(z_1,\bar{z}_1) ]_R [V(z_2,\bar{z}_2)]_R\rangle=&\xi_{1\mu\nu}\xi_{2}{}^{\mu\nu}\left|
\frac{1}{8\pi^2}\left(
\frac{\pi}{\tau_2}+\partial_{\nu}\left[
\frac{\theta_1'(\nu|\tau)}{\theta_1(\nu|\tau)}
\right]
\right)
\right|^2\\
+&
\xi_{1\nu\mu}\xi_{2}{}^{\mu\nu}\left(
\pi\delta^2(z_1-z_2)-\frac{1}{8\pi\tau_2}
\right)^2
+\xi_{1\mu}{}^{\mu}\xi_{2\nu}{}^{\nu}
\left(\frac{1}{8\pi\tau_2}
\right)^2
~.}}
This agrees with \cite{Panda:1987md} except for the fact that there is no $\delta^2(z_1-z_2)$ term in that reference. It also 
agrees with \cite{Abe:1989yb} except for the fact that contrary to that reference there is no $\delta^2(0)$ term in the result above. Finally, \eqref{19} is in agreement with \cite{Minahan:1989cb}. In the following we will use instead the representation of  the Green's function given in (\ref{grsum}), while dropping the zeromode as explained earlier.

Applying the general formula (\ref{ampl}) with renormalized vertices to the case $N=1$, taking 
(\ref{19o}) into account,  we obtain the expression for the one-point  
amplitude:
\eq{\label{a1}
\mathcal{A}_1=g\Lambda\xi_{\mu}{}^{\mu}
~,}
where we have defined:
\eq{\label{ldef}
\Lambda:=C \int_F{\text{d}^2\tau}~\!\tau_2^{-14}\left|\eta(\tau)\right|^{-48}
~,}
and we have rescaled: $g\rightarrow -g\alpha'/8\pi$. Note that $\Lambda$ is divergent, with the divergence coming entirely from the 
tachyon contribution. This can be seen as follows: from the second line in (\ref{ded}) we obtain,
\eq{\label{r}\int_{-\frac12}^{\frac12}
\text{d}\tau_1~\!|\eta(\tau)|^{-48}
=\int_{-\frac12}^{\frac12}\text{d}\tau_1~\!e^{4\pi\tau_2}\big(
1+48\Re q+324\Re q^2+24^2|q|^2+\cdots
\big)=e^{4\pi\tau_2}+24^2+324^2~\!e^{-4\pi\tau_2}+\cdots
~.}
The fundamental domain $F$ 
naturally splits into the upper $\tau_2\geq1$ strip, for which the  $\tau_1$ integration goes from $-\tfrac12$ to $+\tfrac12$ and ensures that only 
the physical on-shell states contribute, and the lower part  for which  $\tau_1$ is excluded between $\pm\sqrt{1-\tau_2^2}$, with $\tfrac{\sqrt{3}}{2}\leq\tau_2<1$.  Contrary to the upper strip, in the lower part of $F$ 
non-physical states contribute as well. In fact, as explained in detail in \cite{Abel:2015oxa}, the dominant contribution in the lower part of $F$ 
comes from nonphysical off-shell tachyonic states. In  \eqref{r} above it is understood that we have restricted to the upper part of $F$.

The exponential term after the last equality in  \eqref{r} is the term responsible for the 
divergence, coming from the $\tau_2\rightarrow\infty$ neighborhood of the integral in (\ref{ldef}). By comparing it to the field theory contribution of a particle of mass $m$ to the one-loop vacuum energy in $D$ dimensions (see e.g. \cite{Polchinski:1998rq}):
\eq{\label{s}\Lambda_m\sim
\int_0^{\infty}\text{d} s s^{-(D+2)/2}e^{-\pi\alpha' m^2s}
~,}
one concludes that the divergence is due to the tachyon.  In the following, we will simply renormalize by hand the value of the one-loop cosmological constant to a finite value, i.e. we will treat  $\Lambda$  as a tunable parameter of the bosonic string model.

Applying  (\ref{ampl}) with renormalized vertices to the case $N=2$, taking (\ref{utp}),(\ref{twopt}) into account, we obtain the expression for the two-point 
amplitude:
\eq{\label{a2}
\mathcal{A}_2=g^2\Lambda\left(
\xi_{1}\xi_{2}
-2\xi_{1\mu\nu}^{(s)}\xi_{2}{}^{(s)\mu\nu}\right)
~,}
where we have defined $\xi_i:=\xi_{i\mu}{}^{\mu}$ and $\xi^{(s)}_{\mu\nu}:=\xi_{(\mu\nu)}$ is the symmetric part of the 
polarization. 
The above expression agrees  with  \cite{Abe:1989yb,Minahan:1989cb}. The authors of \cite{Abe:1989yb,Minahan:1989cb} analyze the corresponding effective action and find mass shifts for the graviton and dilaton but not for the Kalb-Ramond field, in agreement with gauge invariance. The dilaton and graviton ``masses'' are entirely due to the coupling to the vacuum energy.


With a little more effort the analysis can be pursued to $N=3$ at vanishing momentum, in a similar way. Using the identities (\ref{ur}), (\ref{25}), (\ref{tv}) we obtain:
\eq{\label{a3} 
\mathcal{A}_3^{k\to0}=g^3\Lambda\left(
\xi_{1}\xi_{2}\xi_{3}
-2(\xi^{(s)}_{1\mu\nu}\xi_{2}{}^{(s)\mu\nu}\xi_{3} +\mathrm{cyclic})
+8\xi^{(s)}_{1\mu\nu}\xi_{2}{}^{(s)\mu}{}_{\rho}\xi_{3}{}^{(s)\nu\rho}\right)
~,}
where again only the symmetric part of the polarization enters. In particular, we see that there is no coupling of the $B$-field to the 
cosmological constant.  However, the analysis becomes more cumbersome if one wishes to include terms of quadratic or higher order in momenta, or to 
calculate $N$-point amplitudes with $N\geq4$. In the following section we will introduce a diagrammatic technique which facilitates these calculations.

\subsection{Non-derivative couplings}\label{sec:32}

\begin{figure}[tb!]
\begin{center}
\tikzsetnextfilename{node2}
\input{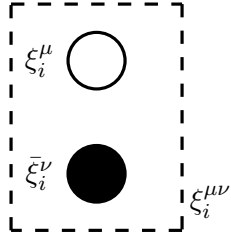}
\end{center}
\caption{The $i$-th  
node represents the polarization $\xi_i^{\mu\nu}$. It consists of one clear and one shaded vertex representing the 
polarizations $\xi_i^{\mu}$ and $\bar{\xi}_i^{\nu}$ respectively.} \label{fig1}
\end{figure}

\begin{figure}[h!]
\begin{center}
\tikzsetnextfilename{admissible_lines}
\input{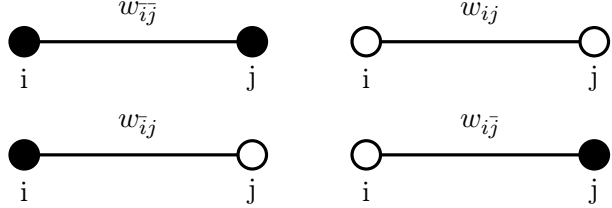}
\end{center}
\caption{The four possible line connections between the vertices of two nodes and their correspondence with the $w$'s.} \label{fig2}
\end{figure}

The following diagrammatics are useful in the calculation of the correlator  in the limit of vanishing external momenta $k_i\rightarrow 0$. To  $S_{1,\dots, N}$ we associate all {\it admissible $N$-graphs}, defined as 
polygons with $N$ {\it nodes}, numbered clockwise from 1 to $N$.  Each 
node corresponds to a polarization $\xi_i^{\mu\nu}$. It consists of a pair of {\it vertices} denoted by a clear and a shaded circle, representing the 
polarizations $\xi_i^{\mu}$, $\bar{\xi}_i^{\nu}$ respectively, as in fig.~\ref{fig1}. 
Each vertex  must be connected to exactly one other vertex by a line, representing the contraction of the corresponding polarizations, together 
with a factor of the form $\partial^2G$ coming from the propagator. 
There are four possible line connections between two nodes, distinguished by the types of vertices they connect, 
each connection being in correspondence with one of the $w$'s as depicted in fig.~\ref{fig2}.

\begin{figure}[h!]
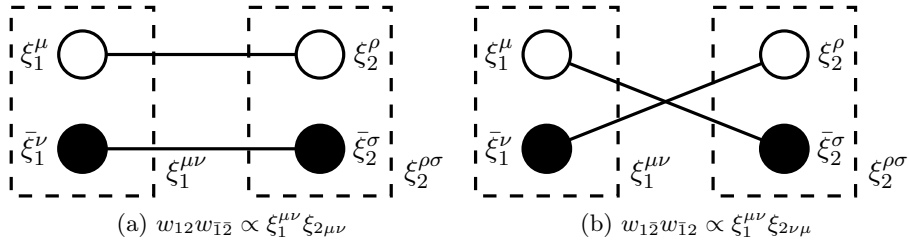

\begin{center}
\subfloat[$w_{12}w_{\bar{1}\bar{2}}\propto \xi_1^{\mu\nu}\xi_{2\mu\nu}$]{
\tikzsetnextfilename{s121p}
\input{s121.tex}}
\subfloat[$w_{1\bar{2}}w_{\bar{1}2}\propto \xi_1^{\mu\nu}\xi_{2\nu\mu}$]{
\tikzsetnextfilename{s122p}
\input{s122.tex}}
\end{center}
\caption{The two admissible graphs corresponding to $S_{1,2}$.} \label{fig3}
\end{figure}

For example $S_{1,2}$ consists of the two admissible graphs depicted in fig.~\ref{fig3}. Each polarization factor listed therein comes multiplied by a term of the form $(\partial^2 G)^2$ which has been suppressed for simplicity. 
This gives,
\eq{\spl{\label{24}
S_{1,2}&=
w_{12}w_{\bar{1}\bar{2}}+w_{\bar{1}2}w_{1\bar{2}}\\
&=\xi_{1\mu\nu}\xi_{2}{}^{\mu\nu}\partial_{1}\partial_{2} G_{12}\partial_{\bar{1}}\partial_{\bar{2}} G_{12}
+\xi_{1\nu\mu}\xi_{2}{}^{\mu\nu}\partial_{\bar{1}}\partial_{2} G_{12}\partial_{{1}}\partial_{\bar{2}} G_{12}
~.
}}
Upon integration over the vertex position both $(\partial^2G)^2$ factors integrate  to $-1/\tau_2(\alpha'/8\pi)^2$, cf.~(\ref{utp}),  
leading to the second term on the right-hand side of (\ref{a2}).

\begin{figure}[tb!]
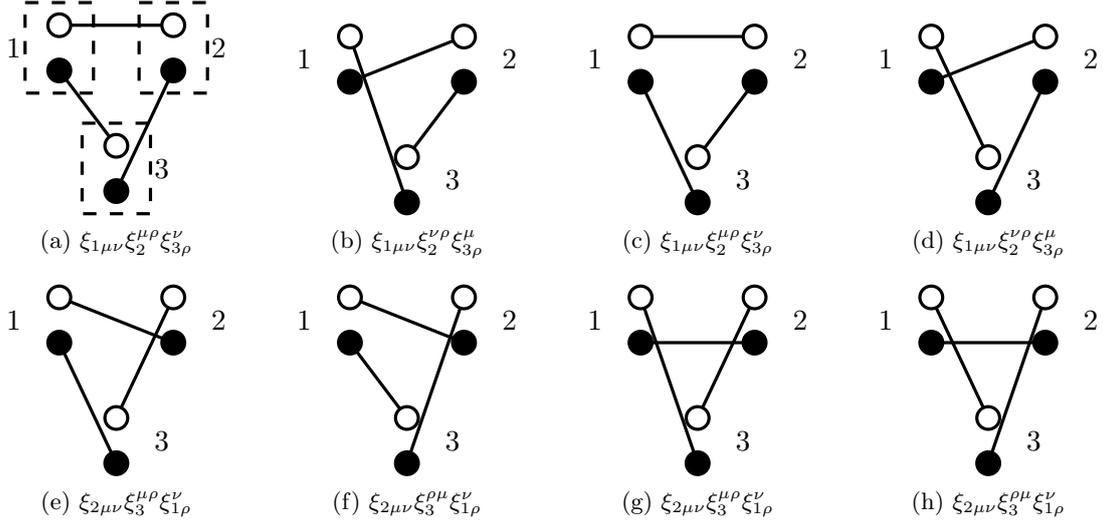

\begin{center}
\begin{tabular}{cccc}
\subfloat[$\xi_{1\mu\nu}\xi_2^{\mu\rho}\xi^{\nu}_{3\rho}$]{
\tikzsetnextfilename{s123ap}
\input{s123a.tex}} 
&
\subfloat[$\xi_{1\mu\nu}\xi_2^{\nu\rho}\xi_{3\rho}^{\mu}$]{
\tikzsetnextfilename{s123bp}
\input{s123b.tex}}
&
\subfloat[$\xi_{1\mu\nu}\xi_2^{\mu\rho}\xi_{3\rho}^{\nu}$]{
\tikzsetnextfilename{s123cp}
\input{s123c.tex}}
&
\subfloat[$\xi_{1\mu\nu}\xi_2^{\nu\rho}\xi^{\mu}_{3\rho}$]{
\tikzsetnextfilename{s123dp}
\input{s123d.tex}}
\\
\subfloat[$\xi_{2\mu\nu}\xi_3^{\mu\rho}\xi^{\nu}_{1\rho}$]{
\tikzsetnextfilename{s123ep}
\input{s123e.tex}}
&
\subfloat[$\xi_{2\mu\nu}\xi_3^{\rho\mu}\xi^{\nu}_{1\rho}$]{
\tikzsetnextfilename{s123fp}
\input{s123f.tex}}
&
\subfloat[$\xi_{2\mu\nu}\xi_3^{\mu\rho}\xi_{1\rho}^{\nu}$]{
\tikzsetnextfilename{s123gp}
\input{s123g.tex}}
&
\subfloat[$\xi_{2\mu\nu}\xi_3^{\rho\mu}\xi_{1\rho}^{\nu}$]{
\tikzsetnextfilename{s123hp}
\input{s123h.tex}}
\end{tabular}
\end{center}
\caption{The eight admissible graphs corresponding to $S_{1,2,3}$. The $(\partial^2 G)^3$ factors have been suppressed.} \label{fig4}
\end{figure}

Similarly $S_{1,2,3}$ consists of the eight admissible graphs depicted in fig.~\ref{fig4}. Each polarization factor listed therein comes multiplied by a term of the schematic 
form $(\partial^2 G)^3$ (omitting the different index structures of the derivatives) which has been suppressed for simplicity. As in the previous example, 
upon integration over the vertex positions all $(\partial^2G)^3$ factors integrate  to $-1/\tau_2(\alpha'/8\pi)^3$, 
leading to the last term on the right-hand side of (\ref{a3}). 

\begin{figure}[h!]
\begin{center}
\tikzsetnextfilename{mixed_node}
\input{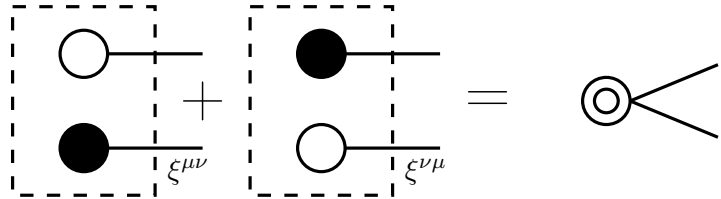}
\end{center}
\caption{The action of exchanging the two vertices within the same node is depicted. Both possibilities are grouped together in a mixed node.} \label{fig5}
\end{figure}

This pattern holds for arbitrary $N$: $S_{1,\dots,N}$ is the sum of all terms of the form (suppressing the different index structures of the derivatives), 
$$\xi_{1\mu_1\nu_1}\dots\xi_{N\mu_N\nu_N}(\partial^2G)^N~,$$ 
where upon integration over the vertex position all $(\partial^2G)^N$ factors  integrate  to the same value, 
\eq{\label{fi}\int \prod_{i=1}^{N-1}\text{d}^2z_i ~(\partial^2G)^N=-\frac{1}{\tau_2}\Big(\frac{\alpha'}{8\pi}\Big)^N
~.}
Moreover the action of exchanging the two vertices within the same node, depicted in fig.~\ref{fig5}, transforms an admissible graph to another admissible graph, and corresponds to the exchange $\xi_i^{\mu\nu}\leftrightarrow\xi_i^{\nu\mu}$. 
This implies that 
only the symmetrized part of the polarization appears in the amplitude in the $k_i\rightarrow 0$ limit. In other words, there is no ``potential'' for the antisymmetric two-form, as 
is of course required for gauge invariance of the amplitude.


To  account for this observation we introduce {\it mixed nodes}, denoted by two cocentric circles, 
grouping together both possibilities, cf.~fig.~\ref{fig5}. 
It can then be seen that each $S_{1,\dots, N}$ is associated with all $N$-polygons with $N$ mixed nodes, each weighted by the combinatorial factor $C_N$ given by,
\eq{C_N=2^{n_1}2^{\frac12 n_2}
~,}
with $n_1:=$the number of nodes whose lines join distinct nodes and $n_2:=$the number of mixed nodes whose lines join the same node. 
Furthermore integration 
over the vertex positions  introduces a multiplicative factor given by (\ref{fi}).

\begin{figure}[h!]
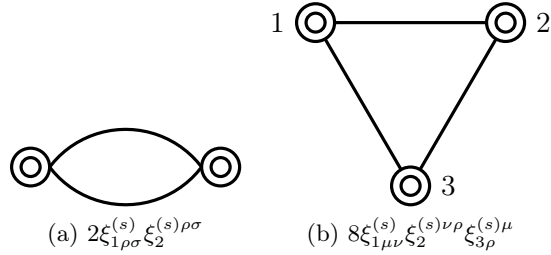

\begin{center}
\subfloat[$2\xi_{1\rho\sigma}^{(s)}\xi_2^{(s)\rho\sigma}$]{
\tikzsetnextfilename{s12_mixedp}
\input{s12_mixed.tex}}
\subfloat[$8\xi_{1\mu\nu}^{(s)}\xi_2^{(s)\nu\rho}\xi_{3\rho}^{(s)\mu}$]{
\tikzsetnextfilename{s123_mixedp}
\input{s123_mixed.tex}}
\end{center}
\caption{The graphs $S_{1,2}$ and $S_{1,2,3}$. We only display the corresponding polarizations with the $(\partial^2 G)^2$ and $(\partial^2 G)^3$ factors suppressed.} \label{fig6}
\end{figure}

Going back to the previous examples: $S_{1,2}$ consists of the first  graph in fig.~\ref{fig6} which gives $n_1=0$, $n_2=1$, $C_2=2$ and $S_{1,2}=2\xi^{(s)}_{1\mu\nu}\xi^{(s)}_{2}{}^{\mu\nu}(\partial^2G)^2$, in accordance with (\ref{24}). 
Similarly $S_{1,2,3}$ is represented by the second graph in fig.~\ref{fig6}. This gives  $n_1=3$, $n_2=0$, $C_3=8$ and $S_{1,2,3}=8\xi^{(s)}_{1\mu\nu}\xi^{(s)}_{2}{}^{\nu\rho}\xi^{(s)}_{3\rho}{}^{\mu}(\partial^3G)^3$. Plugging these into the formula (\ref{ampl}) for the amplitude, taking (\ref{correlt}), (\ref{fi}) into account, reproduces our results for the one-, two- and three-point amplitudes at vanishing external momenta: (\ref{a1}), (\ref{a2}), (\ref{a3}).

\begin{figure}[tb!]
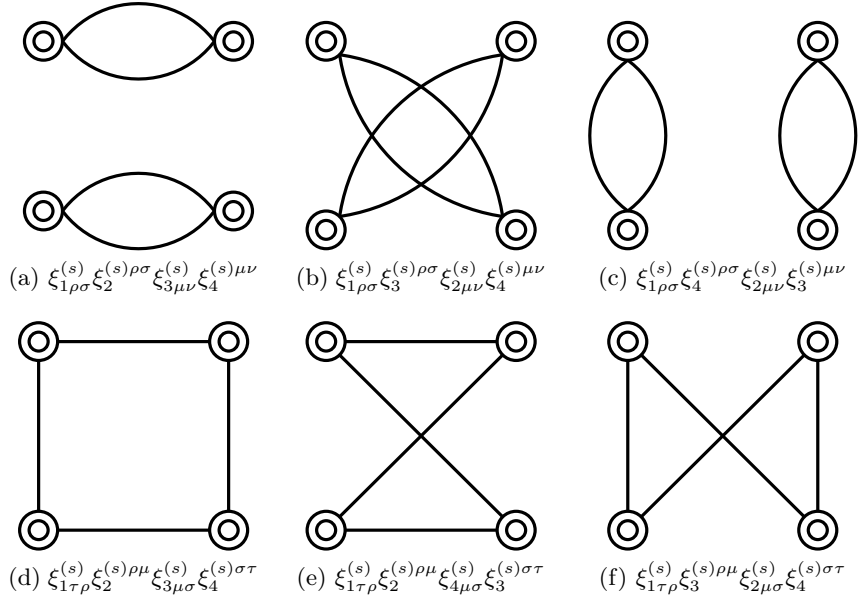

\begin{center}
\begin{tabular}{ccc}
\subfloat[$\xi_{1\rho\sigma}^{(s)}\xi_2^{(s)\rho\sigma}\xi_{3\mu\nu}^{(s)}\xi_4^{(s)\mu\nu}$]{\tikzset{external/remake next}
\tikzsetnextfilename{s1234ap}
\input{s1234a.tex}}
&
\subfloat[$\xi_{1\rho\sigma}^{(s)}\xi_3^{(s)\rho\sigma}\xi_{2\mu\nu}^{(s)}\xi_4^{(s)\mu\nu}$]{
\tikzsetnextfilename{s1234bp}
\input{s1234b.tex}}
&
\subfloat[$\xi_{1\rho\sigma}^{(s)}\xi_4^{(s)\rho\sigma}\xi_{2\mu\nu}^{(s)}\xi_3^{(s)\mu\nu}$]{
\tikzsetnextfilename{s1234cp}
\input{s1234c.tex}}
\\
\subfloat[$\xi_{1\tau\rho}^{(s)}\xi_2^{(s)\rho\mu}\xi_{3\mu\sigma}^{(s)}\xi_4^{(s)\sigma\tau}$]{
\tikzsetnextfilename{s1234dp}
\input{s1234d.tex}}
&
\subfloat[$\xi_{1\tau\rho}^{(s)}\xi_2^{(s)\rho\mu}\xi_{4\mu\sigma}^{(s)}\xi_3^{(s)\sigma\tau}$]{
\tikzsetnextfilename{s1234ep}
\input{s1234e.tex}}
&
\subfloat[$\xi_{1\tau\rho}^{(s)}\xi_3^{(s)\rho\mu}\xi_{2\mu\sigma}^{(s)}\xi_4^{(s)\sigma\tau}$]{
\tikzsetnextfilename{s1234fp}
\input{s1234f.tex}}
\\
\end{tabular}
\end{center}
\caption{The graphs corresponding to $S_{1,2,3,4}$. Those in the first row have 
$C_4=4$, while those in the second row have  $C_4=16$. The $(\partial^2 G)^4$ factors have been suppressed.} \label{fig7}
\end{figure}

The higher-point amplitudes can be evaluated in the same manner. $S_{1,2,3,4}$ is represented by the graphs in fig.~\ref{fig7}. 
The graphs in the first row all have 
 $n_1=0$, $n_2=4$ and $C_4=4$, while those in the second row have  $n_1=4$, $n_2=0$ and $C_4=16$. 
Their total contribution reads,
\eq{\spl{S_{1,2,3,4}=\Big[
4&\big(\xi^{(s)}_{1\rho\sigma}\xi^{(s)}_{2}{}^{\rho\sigma}\xi^{(s)}_{3\mu\nu}\xi^{(s)}_{4}{}^{\mu\nu}
+\xi^{(s)}_{1\rho\sigma}\xi^{(s)}_{3}{}^{\rho\sigma}\xi^{(s)}_{2\mu\nu}\xi^{(s)}_{4}{}^{\mu\nu}
+\xi^{(s)}_{1\rho\sigma}\xi^{(s)}_{4}{}^{\rho\sigma}\xi^{(s)}_{2\mu\nu}\xi^{(s)}_{3}{}^{\mu\nu}
\big)\\
+
16&\big(\xi^{(s)}_{1\mu\nu}\xi^{(s)}_{2}{}^{\nu\rho}\xi^{(s)}_{3\rho\sigma}\xi^{(s)}_{4}{}^{\sigma\mu}
+\xi^{(s)}_{1\mu\nu}\xi^{(s)}_{2}{}^{\nu\rho}\xi^{(s)}_{4\rho\sigma}\xi^{(s)}_{3}{}^{\sigma\mu}
+\xi^{(s)}_{1\mu\nu}\xi^{(s)}_{3}{}^{\nu\rho}\xi^{(s)}_{2\rho\sigma}\xi^{(s)}_{4}{}^{\sigma\mu}
\big)\Big]\times (\partial^4G)^4~,
}}
where $(\partial^2G)^4$ stands schematically for various terms with four propagators and two derivatives on each propagator; the detailed index structure has been suppressed. 
Plugging this into the formula (\ref{ampl}) for the amplitude, taking (\ref{correlt}), (\ref{fi}) into account and rescaling $g\rightarrow-g\alpha'/8\pi$, leads to the following four-point amplitude,

\vfill\break

\eq{\spl{\label{a4} 
\mathcal{A}^{k\rightarrow0}_4=g^4\Lambda&\Big(
\xi_{1}\xi_{2}\xi_{3}\xi_{4}
-2\big[\xi_{1}\xi_{2}{}(\xi_3^{(s)}\cdot\xi_4^{(s)})
+\xi_{1}\xi_{3}{}(\xi_2^{(s)}\cdot\xi_4^{(s)})
+\xi_{1}\xi_{4}{}(\xi_2^{(s)}\cdot\xi_3^{(s)})\\
&~~~~~~~~~~~~~~~~~~~~~~~~~~~~~+\xi_{2}\xi_{3}{}(\xi_1^{(s)}\cdot\xi_4^{(s)})
+\xi_{2}\xi_{4}{}(\xi_1^{(s)}\cdot\xi_3^{(s)})
+\xi_{3}\xi_{4}{}(\xi_1^{(s)}\cdot\xi_2^{(s)})\big]\\
&+8\big[{}(\xi_1^{(s)}\cdot\xi_2^{(s)}\cdot\xi_3^{(s)})\xi_{4} +\mathrm{cyclic}\big]\\
&-16\big[{}(\xi_1^{(s)}\cdot\xi_2^{(s)}\cdot\xi_3^{(s)}\cdot\xi_4^{(s)})
+{}(\xi_1^{(s)}\cdot\xi_2^{(s)}\cdot\xi_4^{(s)}\cdot\xi_3^{(s)})
+{}(\xi_1^{(s)}\cdot\xi_3^{(s)}\cdot\xi_2^{(s)}\cdot\xi_4^{(s)})
\big]\\
&+4\big[{}(\xi_1^{(s)}\cdot\xi_2^{(s)}){}(\xi_3^{(s)}\cdot\xi_4^{(s)})
+{}(\xi_1^{(s)}\cdot\xi_3^{(s)}){}(\xi_2^{(s)}\cdot\xi_4^{(s)})
+{}(\xi_1^{(s)}\cdot\xi_4^{(s)}){}(\xi_2^{(s)}\cdot\xi_3^{(s)})
\big]\Big)
~,}}
where we have adopted a matrix notation: ${}\xi_i^{}\cdot\xi_j^{}:=\xi_{i\mu\nu}^{}\xi_{j}^{\nu\mu}$.

\section{Three-point amplitude, quadratic momentum}\label{sec:3p2k}

Due to kinematic reasons (the fact that $k_i\cdot k_j =k_i\cdot \xi_j=0$ for $i,j=1,2$) there are no two-point couplings with derivatives. 
The three-point two-derivative couplings of three massless particles also vanish identically on-shell due to kinematics. Indeed, imposing momentum conservation  
leads to the relations $k_i\cdot k_j=0$, for all  $i,j=1,2,3$. This then imposes that all three momenta are collinear, which in its turn implies $k_i\cdot \xi_j=0$, for all $i,j=1,2,3$. These 
relations then imply that any Lorentz-invariant three-point amplitude must vanish on shell.   On the other hand, it is known that relaxing the collinearity condition (i.e. formally allowing 
terms of the form $k_i\cdot \xi_j$ to be nonvanishing for $i\neq j$) leads to three-point amplitudes that can be used to correctly reproduce the two-derivative effective action at tree level \cite{Gross:1986mw}. We will assume that this procedure can also be applied to derive the one-loop two-derivative effective action. As we will see in section \ref{sec:eff}  this leads to self-consistent results. A related  recent discussion of off-shell regularization of string amplitudes was given in \cite{Berg:2016wux}, 
based on a certain violation of momentum conservation first introduced in \cite{Minahan:1987ha}.

\begin{figure}[h!]
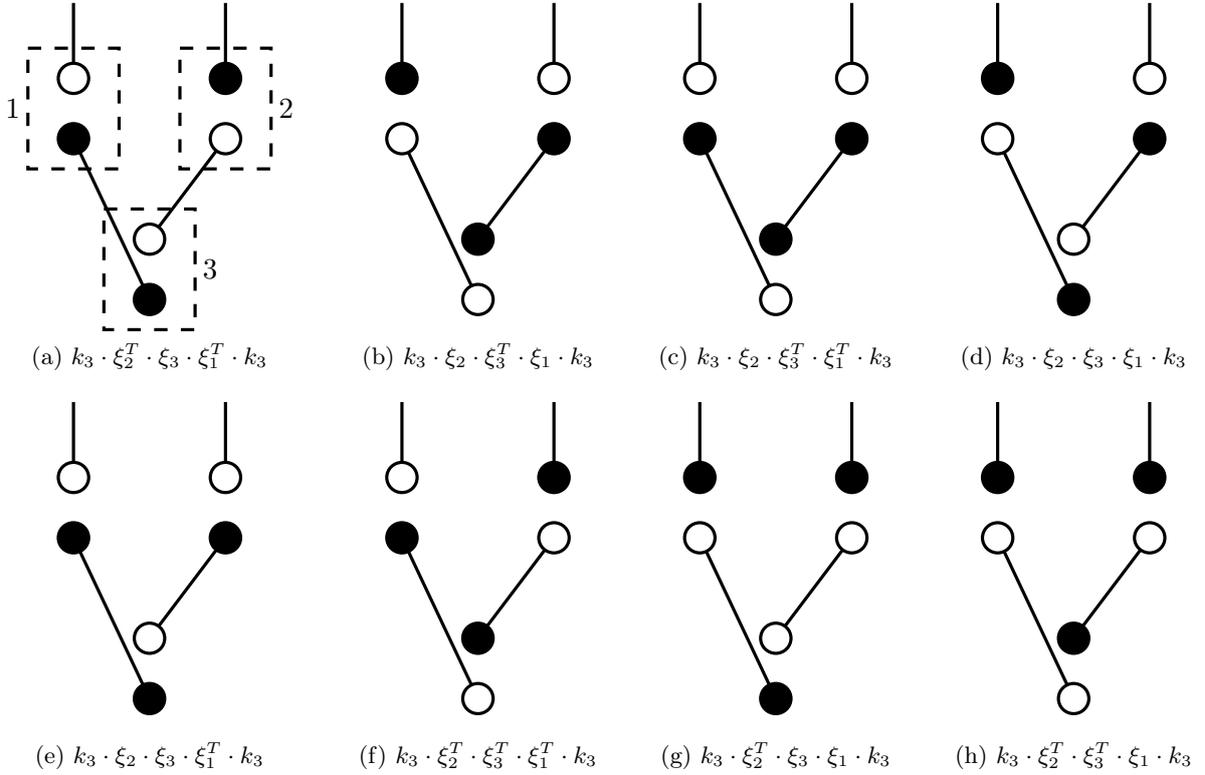

\begin{center}
\begin{tabular}{cccc}
\subfloat[$k_3\cdot \xi_2^T\cdot\xi_3\cdot\xi_1^T\cdot k_3$]{
\tikzsetnextfilename{s123uap}
\input{s123ua.tex}}
&
\subfloat[$k_3\cdot \xi_2\cdot\xi_3^T\cdot\xi_1\cdot k_3$]{
\tikzsetnextfilename{s123ubp}
\input{s123ub.tex}}
&
\subfloat[$k_3\cdot \xi_2\cdot\xi_3^T\cdot\xi_1^T\cdot k_3$]{
\tikzsetnextfilename{s123ucp}
\input{s123uc.tex}}
&
\subfloat[$k_3\cdot \xi_2\cdot\xi_3\cdot\xi_1\cdot k_3$]{
\tikzsetnextfilename{s123udp}
\input{s123ud.tex}}
\\
\subfloat[$k_3\cdot \xi_2\cdot\xi_3\cdot\xi_1^T\cdot k_3$]{
\tikzsetnextfilename{s123uep}
\input{s123ue.tex}}
&
\subfloat[$k_3\cdot \xi_2^T\cdot\xi_3^T\cdot\xi_1^T\cdot k_3$]{
\tikzsetnextfilename{s123ufp}
\input{s123uf.tex}}
&
\subfloat[$k_3\cdot \xi_2^T\cdot\xi_3\cdot\xi_1\cdot k_3$]{
\tikzsetnextfilename{s123ugp}
\input{s123ug.tex}}
&
\subfloat[$k_3\cdot \xi_2^T\cdot\xi_3^T\cdot\xi_1\cdot k_3$]{
\tikzsetnextfilename{s123uhp}
\input{s123uh.tex}}
\\
\end{tabular}
\end{center}
\caption{For each of the graphs of fig.~\ref{fig4}, each of the two vertices connected with a line 
can be replaced by two vertices with open lines to obtain a term bilinear in $u$. This operation leads to 24 graphs,  
obtained from the ones depicted here by cyclic permutations. The corresponding 
polarization factors are listed explicitly.} \label{fig8}
\end{figure}
\begin{figure}[tb!]
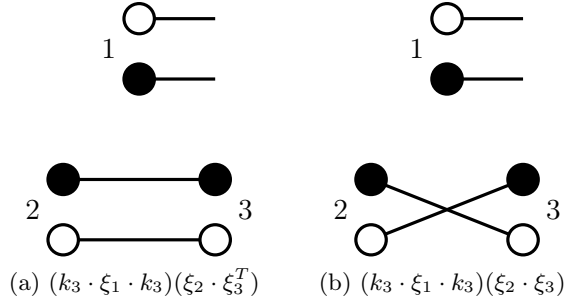

\begin{center}
\begin{tabular}{cc}
\subfloat[$(k_3\cdot \xi_1\cdot k_3)(\xi_2\cdot\xi_3^T)$]{\tikzset{external/remake next}
\tikzsetnextfilename{s123ua2p}
\input{s123ua2.tex}}
&
\subfloat[$(k_3\cdot \xi_1\cdot k_3)(\xi_2\cdot\xi_3)$]{\tikzset{external/remake next}
\tikzsetnextfilename{s123ub2p}
\input{s123ub2.tex}}
\end{tabular}
\end{center}
\caption{Graphs with two open vertices belonging to the same node. The corresponding 
polarization factors are listed explicitly. All graphs of this type are obtained from the ones listed here by cyclic permutations.} \label{fig9}
\end{figure}

To evaluate the 
three-point $\mathcal{O}(\alpha' k^2)$ amplitude we proceed as follows. 
As we see from (\ref{correlt}), (\ref{correlt2}), in the three-point amplitude the only terms quadratic in momenta are those bilinear in $u_i$, $u_{\bar{i}}$. This is because the 
terms $k_i\cdot k_j$ vanish on-shell by momentum conservation and so the exponential term in (\ref{correlt}) becomes trivial. To describe graphically the  terms quadratic in momenta 
we introduce an {\it open vertex}, i.e. a vertex with an open line. This corresponds to $u_i$, $u_{\bar{i}}$, for a clear, shaded vertex respectively. In each of the graphs of fig.~\ref{fig4} we may replace any of the $w$'s with a bilinear in $u$. This corresponds to replacing a pair of vertices connected with a line and  belonging to distinct nodes, by two open vertices. 
When applied to $S_{1,2,3}$, this operation, depicted in fig.~\ref{fig8}, gives 24 graphs. In addition we have graphs with two open vertices belonging to the same node. There is a total of six graphs of this type 
obtained from the ones of fig.~\ref{fig9} by cyclic permutations. 
Let us for example consider the term $u_1u_2w_{\bar{2}3}w_{\bar{1}\bar{3}}$, which corresponds to graph (e) of fig.~\ref{fig8}. It is the sum of four terms each proportional to a polarization factor 
of the form $k_r^{\rho}k_q^{\mu}\xi_{1\mu\nu}\xi_{2\rho\sigma}\xi_3^{\sigma\nu}$, for $q=2,3$, $r=1,3$. 
Setting $q=2$, $r=1$, using similar manipulations as before, after integration over the vertex positions $z_1$, $z_2$ we obtain,


%
\eq{\spl{
-\frac{1}{\pi^2}\Big(
\frac{\alpha'}{8\pi}
\Big)^4&k_1^{\rho}k_2^{\mu}\xi_{1\mu\nu}\xi_{2\rho\sigma}\xi^3_{\sigma\nu}\\
&\times\sum^{\prime}_{
\substack{(m_1,n_1)\\ (m_2,n_2)\\(m_1,n_1)\neq(m_2,n_2)}
}\frac{(m_1-n_1\bar{\tau})(m_2-n_2\bar{\tau})[(m_1-m_2)-(n_1-n_2){\tau}]^2}{|m_1-n_1{\tau}|^2|m_2-n_2{\tau}|^2|(m_1-m_2)-(n_1-n_2){\tau}|^2}
~,
}}

where the prime above the sum symbol indicates that $(m_1,n_1)$, $(m_2,n_2)\neq(0,0)$. The sum in the second line above can be evaluated as follows. 
Expanding the square in the last term,
\eq{\spl{\label{interm1}
&\sum^{\prime}_{
\substack{(m_1,n_1)\\ (m_2,n_2)\\(m_1,n_1)\neq(m_2,n_2)}
}\frac{(m_1-n_1\bar{\tau})(m_2-n_2\bar{\tau})[(m_1-m_2)-(n_1-n_2){\tau}]^2}{|m_1-n_1{\tau}|^2|m_2-n_2{\tau}|^2|(m_1-m_2)-(n_1-n_2){\tau}|^2}\\
=&\sum^{\prime}_{
\substack{(m_1,n_1)\\ (m_2,n_2)\\(m_1,n_1)\neq(m_2,n_2)}
}\frac{2(m_1-n_1\bar{\tau})(m_2-n_2{\tau})}{|m_1-n_1{\tau}|^2|(m_1-m_2)-(n_1-n_2){\tau}|^2}
-\frac{2}{|(m_1-m_2)-(n_1-n_2){\tau}|^2}\\
=&\frac{2}{\tau_2}E_1(\tau)
~.
}}
\vfill\break

The last equality can be seen as follows. First note that,
\eq{\spl{\label{interm2}
&\sum^{\prime}_{
\substack{(m_1,n_1)\\ (m_2,n_2)\\(m_1,n_1)\neq(m_2,n_2)}
}\frac{1}{|(m_1-m_2)-(n_1-n_2){\tau}|^2}\\
=
&\sum^{\prime}_{
(m_1,n_1)
}
\left(
\sum_{
\substack{(m_2,n_2)\\(m_2,n_2)\neq(m_1,n_1)}
}
\frac{1}{|(m_1-m_2)-(n_1-n_2){\tau}|^2}
-\frac{1}{|(m_1-0)-(n_1-0){\tau}|^2}
\right)\\
=
&\sum^{\prime}_{
(m_1,n_1)
}\frac{1}{|m_1-n_1{\tau}|^2} 
\sum^{\prime}_{
(m_1,n_1)
}1-\sum^{\prime}_{
(m_1,n_1)
}\frac{1}{|m_1-n_1{\tau}|^2} =-\frac{2}{\tau_2}E_1(\tau)
~,
}}
where we took (\ref{e1}), (\ref{e2}) into account. In order to evaluate the first term in the second line of (\ref{interm1}) we note the following identity:


%
%
%
\eq{\spl{\label{interm3}
&\sum^{\prime}_{
\substack{(m_1,n_1)\\ (m_2,n_2)\\(m_1,n_1)\neq(m_2,n_2)}
}
\frac{(m_1-n_1{\tau})^2(m_2-n_2\bar{\tau})^2}{|m_1-n_1{\tau}|^2|m_2-n_2{\tau}|^2|(m_1-m_2)-(n_1-n_2){\tau}|^2}\\
&=\sum^{\prime}_{
\substack{(m_1,n_1)\\ (m_2,n_2)\\(m_1,n_1)\neq(m_2,n_2)}
}
\frac{(m_1-n_1{\tau})^2[(m_1-m_2)-(n_1-n_2)\bar{\tau}]^2}{|m_1-n_1{\tau}|^2|m_2-n_2{\tau}|^2|(m_1-m_2)-(n_1-n_2){\tau}|^2}
\\
&=\sum^{\prime}_{
\substack{(m_1,n_1)\\ (m_2,n_2)\\(m_1,n_1)\neq(m_2,n_2)}
}\Big(
\frac{(m_1-n_1{\tau})^2(m_2-n_2\bar{\tau})^2}{|m_1-n_1{\tau}|^2|m_2-n_2{\tau}|^2|(m_1-m_2)-(n_1-n_2){\tau}|^2}\\
&~~~~~~~~~~~~~~~+\frac{|m_1-n_1{\tau}|^2}{|m_2-n_2{\tau}|^2|(m_1-m_2)-(n_1-n_2){\tau}|^2}
-\frac{2(m_1-n_1{\tau})(m_2-n_2\bar{\tau})}{|m_1-n_1{\tau}|^2|(m_1-m_2)-(n_1-n_2){\tau}|^2}\Big)~,
}}
which implies,
\eq{\spl{\label{interm4}
\sum^{\prime}_{
\substack{(m_1,n_1)\\ (m_2,n_2)\\(m_1,n_1)\neq(m_2,n_2)}
}&
\frac{2(m_1-n_1{\tau})(m_2-n_2\bar{\tau})}{|m_1-n_1{\tau}|^2|(m_1-m_2)-(n_1-n_2){\tau}|^2}\\
&=\sum^{\prime}_{
\substack{(m_1,n_1)\\ (m_2,n_2)\\(m_1,n_1)\neq(m_2,n_2)}
}\frac{|m_1-n_1{\tau}|^2}{|m_2-n_2{\tau}|^2|(m_1-m_2)-(n_1-n_2){\tau}|^2}
=-\frac{2}{\tau_2}E_1(\tau)
~,
}}
where in the last equality we made use of a  result in \cite{DHoker:2015gmr}. Indeed the sum in the second line is a special case of the infinite sums $C_{a,b,-1}$ of \cite{DHoker:2015gmr} which were shown therein to satisfy,
\eq{C_{a,b,-1}=E_{a-1}E_b+E_{b-1}E_a~,}
where $E_a:=\pi^{-a}E(\tau,a)$ and $C_{1,1,-1}$ equals $1/\pi$ times the sum in second line of (\ref{interm4}). Setting $a=b=1$ in the above and taking (\ref{e2}) into account leads 
to the last equality in (\ref{interm4}). Inserting (\ref{interm2}), (\ref{interm4}) in the second line of (\ref{interm1}) leads to the last equality therein. 

In conlusion, the  $q=2$, $r=1$ term of $u_1u_2w_{\bar{2}3}w_{\bar{1}\bar{3}}$ gives:
\eq{\label{reco}
-\frac{2}{\tau_2\pi^2}\Big(
\frac{\alpha'}{8\pi}
\Big)^4 k_1^{\rho}k_2^{\mu}\xi_{1\mu\nu}\xi_{2\rho\sigma}\xi_3^{\sigma\nu} E_1(\tau)
~.
}
Alternatively, this term may be computed integrating by parts the $\bar{\partial}_{z_1}$ derivative in $w_{\bar{1}\bar{3}}$. This 
leads to the following expression:
\eq{\spl{\label{reco2}
\frac{2}{\pi^2}\Big(
\frac{\alpha'}{8\pi}
\Big)^4&k_1^{\rho}k_2^{\mu}\xi_{1\mu\nu}\xi_{2\rho\sigma}\xi_3^{\sigma\nu}\times\\
&\sum^{\prime}_{
\substack{(m_1,n_1)\\ (m_2,n_2)\\(m_1,n_1)\neq(m_2,n_2)}
}
\frac{(m_1-n_1\bar{\tau})[(m_1-m_2)-(n_1-n_2){\tau}]}{|m_1-n_1\bar{\tau}|^2|(m_1-m_2)-(n_1-n_2){\tau}|^2}
~,
}}
after integration over the vertex positions. 
The sum above can be computed as follows,
\eq{ \spl{\label{interm5}
\sum^{\prime}_{
\substack{(m_1,n_1)\\ (m_2,n_2)\\(m_1,n_1)\neq(m_2,n_2)}
}\!\!\!\!\!\!\!\!
\frac{(m_1-n_1\bar{\tau})[(m_1-m_2)-(n_1-n_2){\tau}]}{|m_1-n_1\bar{\tau}|^2|(m_1-m_2)-(n_1-n_2){\tau}|^2}&=\\
\!\!\!
\sum^{\prime}_{
\substack{(m_1,n_1)\\ (m_2,n_2)\\(m_1,n_1)\neq(m_2,n_2)}
}\!\!\!\!\!\!\!\!
\frac{(m_1-n_1\bar{\tau})(m_2-n_2{\tau})}{|m_1-n_1{\tau}|^2|m_2-n_2{\tau}|^2}
&=
-\frac{1}{\tau_2}E_1(\tau)
~.
}}
where in the last equality we took into account that, without the restriction $(m_1,n_1)\neq(m_2,n_2)$, the sum would vanish as a consequence of its antisymmetry 
under the exchange $(m_i,n_i)\leftrightarrow-(m_i,n_i)$ for $i=1$ or $i=2$. Moreover the contribution of the terms with $(m_1,n_1)=(m_2,n_2)$ can easily be computed, leading to the result above.  
Inserting \eqref{interm5} in (\ref{reco2}) we then recover (\ref{reco}).

The  remaining contributions to $u_1u_2w_{\bar{2}3}w_{\bar{1}\bar{3}}$ from the other values of  $q$, $r$ can similarly be calculated using the results for the infinite sums derived above: the contribution from $q=3$, $r=1$ cancels the one in (\ref{reco}); the contribution from $q=2$, $r=3$ equals one-half that of (\ref{reco}), while
 the contribution from $q=3$, $r=3$ vanishes. Summing up all contributions for graph (e) we obtain,
\eq{\label{type8p}
\frac{1}{\tau_2\pi^2}\Big(
\frac{\alpha'}{8\pi}
\Big)^4 E_1(\tau)\times k_3\cdot\xi^{}_{2}\cdot\xi^{}_{3}\cdot\xi^{T}_{1} \cdot k_3
~,
}
where we have adopted matrix notation for the polarizations. 
The numerical factor above turns out to be equal to the numerical factor multiplying the polarization of graphs (c)-(h) of fig.~(\ref{fig8}), while the 
numerical factor of the graphs (a) and (b) is $(-5\times)$ the above.   
The total sum of all graphs of the type depicted in fig.~(\ref{fig8}) can thus be put in the form,
\eq{\label{type8}
\frac{1}{\tau_2\pi^2}\Big(
\frac{\alpha'}{8\pi}
\Big)^4 E_1(\tau)\times k_3\cdot\left(
8\xi^{(s)}_{2}\cdot\xi^{(s)}_{3}\cdot\xi^{(s)}_{1} 
-6 \xi^{}_{2}\cdot\xi^{T}_{3}\cdot\xi^{}_{1}  
-6\xi^{T}_{2}\cdot\xi^{}_{3}\cdot\xi^{T}_{1} 
\right)\cdot k_3+\mathrm{cyclic}
~.
}
The calculation of the contributions of the graphs depicted in fig.~(\ref{fig9}) proceeds similarly. 
Let us first consider the term $u_1u_{\bar{1}}w_{23}w_{\bar{2}\bar{3}}$: 
it is the sum of four terms each proportional to a polarization factor 
of the form $k_q^{\mu}k_r^{\nu}\xi_{1\mu\nu}\xi_{2\rho\sigma}\xi_3^{\rho\sigma}$, for $q,r=2,3$. 
Setting $q=2$, $r=3$, after integration over the vertex positions $z_1$, $z_2$,  we obtain,
\eq{\spl{\label{intem6}
-\frac{1}{\pi^2}\Big(
\frac{\alpha'}{8\pi}
\Big)^4&k_2^{\mu}k_3^{\nu}\xi_{1\mu\nu}\xi_{2\rho\sigma}\xi_3^{\rho\sigma}\\
&\times\sum^{\prime}_{
\substack{(m_1,n_1)\\ (m_2,n_2)\\(m_1,n_1)\neq(m_2,n_2)}
}\frac{(m_2-n_2\bar{\tau})^2[(m_1-m_2)-(n_1-n_2){\tau}]^2}{|m_1-n_1{\tau}|^2|m_2-n_2{\tau}|^2|(m_1-m_2)-(n_1-n_2){\tau}|^2}~.
}}
The sum above can be computed indirectly by integrating by parts the $\partial_{z_2}$ derivative before performing the integration over the vertex positions 
and using some of the previous results. This gives,
\eq{\spl{\label{interm7}
&\sum^{\prime}_{
\substack{(m_1,n_1)\\ (m_2,n_2)\\(m_1,n_1)\neq(m_2,n_2)}
}
\frac{(m_1-n_1{\tau})^2(m_2-n_2\bar{\tau})^2}{|m_1-n_1{\tau}|^2|m_2-n_2{\tau}|^2|(m_1-m_2)-(n_1-n_2){\tau}|^2}\\
&=\sum^{\prime}_{
\substack{(m_1,n_1)\\ (m_2,n_2)\\(m_1,n_1)\neq(m_2,n_2)}
}\frac{[(m_1-m_2)-(n_1-n_2){\tau}]^2(m_2-n_2\bar{\tau})^2}{|m_1-n_1{\tau}|^2|m_2-n_2{\tau}|^2|(m_1-m_2)-(n_1-n_2){\tau}|^2}
=
\frac{1}{\tau_2}E_1(\tau)
~,}}
where the first equality is obtained by a change of variables, $m_1\rightarrow m_2-m_1$, $n_1\rightarrow n_2-n_1$. 
The contribution from all other values of $q$, $r$ turns out to be equal to the above, giving in total,
\eq{\label{type9kk}
-\frac{4}{\tau_2\pi^2}\Big(
\frac{\alpha'}{8\pi}
\Big)^4 E_1(\tau)\times  (k_2\cdot\xi^{}_{1}\cdot k_3) ~(\xi^{}_{2}\cdot\xi^{T}_{3}) 
~,
}
for the first graph of fig.~(\ref{fig9}). Similarly, the second graph gives a total contribution of,
\eq{\label{type9kkkf}
\frac{2}{\tau_2\pi^2}\Big(
\frac{\alpha'}{8\pi}
\Big)^4 E_1(\tau)\times  (k_2\cdot\xi^{}_{1}\cdot k_3) ~(\xi^{}_{2}\cdot\xi^{}_{3})
~.
}
The sum of all graphs of the type depicted in fig.~(\ref{fig9}) can thus be put in the form,
\eq{\label{type9}
\frac{1}{\tau_2\pi^2}\Big(
\frac{\alpha'}{8\pi}
\Big)^4 E_1(\tau)\times  
\left(
-2(k_2\cdot\xi^{(s)}_{1}\cdot k_3) ~(\xi^{(s)}_{2}\cdot\xi^{(s)}_{3}) 
+6(k_2\cdot\xi^{(s)}_{1}\cdot k_3) ~(\xi^{(a)}_{2}\cdot\xi^{(a)}_{3}) 
+\mathrm{cyclic}
\right)
~.
}
In addition to (\ref{type8}), (\ref{type9}) we have the contribution from terms of the form $\xi_3S_{1,2}$ and cyclic permutations thereof, cf.~(\ref{correlt}). Keeping the 
quadratic-momentum coupling in (\ref{correlt2}) we obtain,
\eq{\label{56ty}S_{1,2}=
u_{\bar{1}}u_{\bar{2}}w_{{12}}+
u_{\bar{1}}u_{{2}}w_{1\bar{2}}+
u_{{1}}u_{\bar{2}}w_{\bar{1}2}+
u_{{1}}u_{{2}}w_{\bar{1}\bar{2}}
~.}
Let us first consider the term $u_{\bar{1}}u_{\bar{2}}w_{{12}}$. It is the sum of four terms each proportional to a polarization factor 
of the form $k_q^{\nu}k_r^{\rho}\xi_{1\mu\nu}\xi_2^{\mu}{}_{\rho}\xi_3$, for $q=2,3$, $r=1,3$. 
Setting $q=2$, $r=1$,  integrating over the vertex positions $z_1$, $z_2$ and using (\ref{interm1}) we obtain,
\eq{\label{dfg}-\frac{2}{\tau_2\pi^2}\Big(
\frac{\alpha'}{8\pi}
\Big)^4 E_1(\tau)\times
k_2^{\nu}k_1^{\rho}\xi_{1\mu\nu}\xi_2^{\mu}{}_{\rho}\xi_3
~.
}
Alternatively we may arrive at the same result by first integrating the $\bar{\partial}_{z_1}$ derivative by parts and using the identities in \eqref{interm2}, \eqref{interm4},
\eq{\label{intermtrt2}
\sum^{\prime}_{
\substack{(m_1,n_1)\\ (m_2,n_2)\\(m_1,n_1)\neq(m_2,n_2)}
}\frac{(m_1-n_1\bar{\tau})[(m_1-m_2)-(n_1-n_2){\tau}]}{|m_1-n_1{\tau}|^2|m_2-n_2{\tau}|^2}
=\frac12\times \!\!\!\!\!\!\!\!\sum^{\prime}_{
\substack{(m_1,n_1)\\ (m_2,n_2)\\(m_1,n_1)\neq(m_2,n_2)}
}\frac{1}{|m_1-n_1{\tau}|^2}
=-\frac{1}{\tau_2}E_1(\tau)
~.
}
{}Furthermore, the terms with $q=2$, $r=3$ and   $q=3$, $r=1$ can be seen to vanish, while the term with $q=r=3$ gives $(-1/2)$ the contribution of (\ref{dfg}).

The second term of (\ref{56ty}), $u_{\bar{1}}u_{{2}}w_{1\bar{2}}$, can be calculated with similar manipulations, giving minus the contribution of (\ref{dfg}), while the 
third and fourth terms in (\ref{56ty}) are obtained from the second and first respectively by exchanging $\xi_i^{\mu\nu}\leftrightarrow\xi_i^{\nu\mu}$ for $i=1,2$. The total 
contribution of the terms of the form $\xi_iS_{\!\!j,k}$ thus reads,
\eq{\label{dfg2}
\frac{1}{\tau_2\pi^2}\Big(
\frac{\alpha'}{8\pi}
\Big)^4 E_1(\tau)\times\left(
k_3\cdot \big(
2\xi_1^{(s)}\cdot \xi_{2}^{(s)} 
-6\xi_1^{(a)} \cdot\xi_2^{(a)}
\big)\cdot k_3~\!\xi_3
+\mathrm{cyclic}\right)
~.
}
The three-point quadratic-momentum amplitude, before integration over the torus modulus,  is the sum of (\ref{type8}), (\ref{type9}), (\ref{dfg2}). We distinguish the following three cases:
 
$\bullet$ {\it Odd number of antisymmetric polarizations}.  Eqs.~(\ref{type8}), (\ref{type9}), (\ref{dfg2}) can be seen to vanish identically in this case. 

$\bullet$ {\it Three symmetric polarizations}.  The total sum is,
\eq{\spl{\label{3s}
-&\frac{2}{\tau_2\pi^2}\Big(
\frac{\alpha'}{8\pi}
\Big)^4 E_1(\tau)\times\\
\Big(&
-(k_3\cdot  
\xi_1^{(s)}\cdot \xi_{2}^{(s)}  \cdot k_3)~\!\xi_3
+(k_2\cdot\xi^{(s)}_{1}\cdot k_3) ~(\xi^{(s)}_{2}\cdot\xi^{(s)}_{3}) 
+2(k_3\cdot 
\xi^{(s)}_{2}\cdot\xi^{(s)}_{3}\cdot\xi^{(s)}_{1} \cdot k_3)
+\mathrm{cyclic}\Big)
~.
}}
$\bullet$ {\it Two  antisymmetric polarizations}.  The total sum is,
\eq{\spl{\label{2b}
-&\frac{6}{\tau_2\pi^2}\Big(
\frac{\alpha'}{8\pi}
\Big)^4 E_1(\tau)\times\\
\Big(&2
k_1\cdot 
\xi^{(s)}_{2}\cdot\xi^{(a)}_{3}\cdot\xi^{(a)}_{1} \cdot k_3
+2
k_1\cdot 
\xi^{(a)}_{2}\cdot\xi^{(s)}_{3}\cdot\xi^{(a)}_{1} \cdot k_2
+2
k_3\cdot 
\xi^{(a)}_{2}\cdot\xi^{(a)}_{3}\cdot\xi^{(s)}_{1} \cdot k_2\\
&~~~~~~~~~~~~~~~~~~~~~~~~~~-(k_2\cdot\xi^{(s)}_{1}\cdot k_3) ~(\xi^{(a)}_{2}\cdot\xi^{(a)}_{3}) 
-(k_3\cdot\xi_1^{(a)} \cdot\xi_2^{(a)}\cdot k_3)~\!\xi_3
+\mathrm{cyclic}\Big)
~.
}}
The amplitude  before integration over the torus modulus is proportional to $E_1(\tau)$, which has a pole divergence. 
We will renormalize as follows,\footnote{As can be seen from (\ref{ldef}), one could arrive at the same renormalization by setting $-\tfrac{1}{8\pi^3}E_1(\tau)
\rightarrow c$ and rescaling $g\rightarrow -g\alpha'/8\pi$ as usual. However this would be  more restrictive than  \eqref{r2c}.}
\eq{\label{r2c}
Cg^3 \int_F{\text{d}^2\tau}~\!\tau_2^{-14}\left|\eta(\tau)\right|^{-48}
\frac{1}{\pi^2}\Big(
\frac{\alpha'}{8\pi}
\Big)^4 E_1(\tau)\rightarrow  c g^3  \alpha' \Lambda
~.}
The tachyonic divergence is included in $\Lambda$, cf.~(\ref{ldef}); the renormalization 
(\ref{r2c}) amounts to removing the IR divergences coming from dilaton or off-shell graviton propagation along long-tube degenerations of the torus, see e.g.~\cite{Tseytlin:1988mw}, of the kind depicted in figs.~\ref{urfig1}, \ref{urfig2}.

It is instructive to compare the structure of the one-loop amplitude (\ref{3s}), (\ref{2b}) to that of the corresponding tree-level amplitude,
\eq{\spl{\label{atree}
\mathcal{A}^{\text{tree}}_3&\propto
(k_2\cdot\xi^{(s)}_{1}\cdot k_3) ~(\xi^{(s)}_{2}\cdot\xi^{(s)}_{3}) 
+2(k_3\cdot 
\xi^{(s)}_{2}\cdot\xi^{(s)}_{3}\cdot\xi^{(s)}_{1} \cdot k_3)\\
&+
2
k_1\cdot 
\xi^{(s)}_{2}\cdot\xi^{(a)}_{3}\cdot\xi^{(a)}_{1} \cdot k_3
+2
k_1\cdot 
\xi^{(a)}_{2}\cdot\xi^{(s)}_{3}\cdot\xi^{(a)}_{1} \cdot k_2
+2
k_3\cdot 
\xi^{(a)}_{2}\cdot\xi^{(a)}_{3}\cdot\xi^{(s)}_{1} \cdot k_2\\
&-(k_2\cdot\xi^{(s)}_{1}\cdot k_3) ~(\xi^{(a)}_{2}\cdot\xi^{(a)}_{3}) 
+\mathrm{cyclic}~.
}}
We see that the first line corresponds precisely to the structure in (\ref{3s}), except for the trace term (proportional to $\xi\equiv\xi^{\mu}{}_{\mu}$) which is absent at tree level. 
Similarly the last two lines reproduce the structure in (\ref{2b}), up to  the trace term. 
This is not surprising in view of the fact that, as we shall see in section \ref{sec:eff}, the structures in  (\ref{3s}), (\ref{2b}) are completely determined (up to an overall coefficient) by 
the form of the Einstein term and the three-form kinetic term respectively. 
One difference from the one-loop amplitude is that the relative factor  
between the two structures in (\ref{3s}), (\ref{2b}) is three times the one at tree level. This is also reflected in the relative coefficients of the Einstein and  three-form kinetic terms 
in the one-loop effective action, cf.~eq.~(\ref{sefffinresc1}).

\section{Effective two-derivative action}\label{sec:eff}

In order to read off the effective action from the amplitude, 
we proceed  to expand the momentum-space 
polarization tensor as in \cite{Gross:1986mw},
\eq{\label{31}\xi_{\mu\nu}=
h_{\mu\nu}(k)+b_{\mu\nu}(k)
+\tfrac{1}{\sqrt{D-2}}\phi(k)\bar{\eta}_{\mu\nu}(k)
~,}
where $h_{\mu\nu}$ is symmetric transverse ($k^{\mu}h_{\mu\nu}=0$) and traceless ($h_{\mu}{}^{\mu}=0$), $b_{\mu\nu}$ is antisymmetric transverse, and 
\eq{\label{etab}\bar{\eta}_{\mu\nu}(k):=\eta-k_{\mu}\bar{k}_{\nu}-k_{\nu}\bar{k}_{\mu}~,}
where $\bar{k}_{\mu}$ is an arbitrary vector obeying, 
\eq{\label{auxr}k\cdot\bar{k}=1~,~~~ \bar{k}\cdot\bar{k}=0~.} 
This definition ensures that $\bar{\eta}_{\mu\nu}$ is 
symmetric transverse. The tensor $h_{\mu\nu}$ and the scalar $\phi$ will be identified with the graviton and the Einstein-frame dilaton respectively. 
%
%
%
%
%

We would now like to use the string amplitudes 
to reconstruct the one-loop effective action. To that end we first postulate canonical kinetic terms for the graviton $h_{\mu\nu}$ and the Einstein-frame dilaton $\phi$, see e.g. \cite{Gross:1986mw}:
\eq{\label{ae}
S_k=\int\text{d}^Dx\sqrt{G}\big(\frac{1}{2\kappa^2}R-\frac12 (\partial\phi)^2
-\frac16 H^2e^{-\frac{4\kappa}{\sqrt{D-2}}~\!\phi}\big)
~,}
where $H_{\mu\nu\rho}:=3\partial_{[\mu}b_{\nu\rho]}$ and $G_{\mu\nu}:=\eta_{\mu\nu}+2\kappa h_{\mu\nu}$ is the Einstein-frame metric expanded around flat space; $\kappa$ is the gravitational coupling, 
and we have covariantized the graviton kinetic term. In the string frame the action (\ref{ae}) takes the form,
\eq{\label{as}
S_k=\frac{1}{2\kappa^2}\int\text{d}^Dx\sqrt{{G}'}e^{-2\varphi}\big(R({G}')+4 (\partial\varphi)^2
-\frac16 H^2
\big)
~,}
where the string-frame metric ${G}'$  and the string-frame dilaton $\varphi$ are given by,
\eq{\label{sf}{G}_{\mu\nu}=e^{-\frac{4}{D-2}\varphi}{G}'_{\mu\nu}~;~~~
\phi=\tfrac{2}{\kappa\sqrt{D-2}}~\!\varphi
~.}
The effective action at one-loop order in the string coupling includes a cosmological constant and corrections 
to the two-derivative kinetic terms. Its general form reads,
\eq{\label{ac}
S_{1-loop}=
\int\text{d}^Dx\sqrt{G}\Big\{
e^{\kappa\sqrt{D-2}~\!\!\phi}\big(
c_1 \frac{1}{2\kappa^2}R-c_2\frac12 (\partial\phi)^2
-c_3\frac16 H^2e^{-\frac{4\kappa}{\sqrt{D-2}}~\!\!\phi}\big)
+
{c_4}e^{\frac{\kappa D}{\sqrt{D-2}}~\!\!\phi}
\Big\}
~,}
where we took into account the expected weight in the string frame, $e^{2(l-1)\varphi}$, for the couplings generated at $l$-loop order. The coefficients 
$c_1, \dots, c_4$ will be determined by comparison to the string amplitudes.


{\it Comparison of (\ref{ac}) with the $N$-point amplitudes at vanishing external  momenta}:

Taking into account the expansion,
\eq{
\spl{\label{sre}\sqrt{1 +2\kappa A}=1&+\kappa \mathrm{tr}A+\frac12 \kappa^2\big[(\mathrm{tr}A)^2
-2\mathrm{tr}(A^2)\big]
+\frac{1}{3!} \kappa^3\big[(\mathrm{tr}A)^3-6\mathrm{tr}A\mathrm{tr}(A^2)
+8\mathrm{tr}(A^3)\big]
\\
&+\frac{1}{4!} \kappa^4\big[(\mathrm{tr}A)^4
+32\mathrm{tr}A\mathrm{tr}(A^3)
-12(\mathrm{tr}A)^2\mathrm{tr}(A^2)+12\big(\mathrm{tr}(A^2)\big)^2
-48\mathrm{tr}(A^4)\big]
+\mathcal{O}(\kappa^5)
~,
}}
we see that 
the $N$-point gravitational amplitudes  at vanishing external  momenta, (\ref{a1}), (\ref{a2}), (\ref{a3}), (\ref{a4}), $N=1,\dots, 4$, are consistent with the term,
\eq{\label{acalt}
\Lambda\sqrt{{\eta_{\mu\nu}+2gh_{\mu\nu}}}
~,}
in the 
effective Lagrangian
where we have used (\ref{31}). 
Comparing with (\ref{ac})  leads to the following identifications,
\eq{\label{310}
g=\kappa~;~~~c_4=\Lambda
~.}
The coupling of the dilaton to the cosmological constant in (\ref{ac}) 
is also consistent with the $N$-point amplitudes (\ref{a1}), (\ref{a2}), (\ref{a3}), (\ref{a4}) provided we use the identifications: 
$\xi^{(s)}_{\mu\nu}\rightarrow h_{\mu\nu}$ and $\xi\equiv\xi^{\mu}{}_{\mu}\rightarrow D~\!\phi/\sqrt{D-2}$, the trace of the polarization tensor at vanishing momentum, cf.~(\ref{31}).

{\it Comparison of (\ref{ac}) with the three-point amplitudes at quadratic momentum}:

We will now use  the three-point amplitude derived earlier in order to read off the two-derivative effective action at  one-loop.

$\bullet$  The $hhh$ coupling

The one-loop correction to the graviton kinetic term (which is covariantized to the scalar curvature) can be derived from the $hhh$ coupling in (\ref{3s}), 
i.e. substituting $\xi_{\mu\nu}^{(s)}\rightarrow h_{\mu\nu}$ therein and keeping only the last two terms; the ``trace'' term (proportional to $\xi_3$) and its cyclic permutations do not contribute to the coupling. 
To compare this to  (\ref{ac}), we expand $G_{\mu\nu}=\eta_{\mu\nu}+2\kappa h_{\mu\nu}$ therein and keep the term cubic in $h$.  After passing to momentum space, this gives,
\eq{
-2\kappa c_1~\Big(
(k_2\cdot h_{1}\cdot k_3) ~(h_{2}\cdot h_{3}) 
+2(k_2\cdot 
h_{1}\cdot  h_{2}\cdot h_{3} \cdot k_2)
+\mathrm{cyclic}\Big)
~.}
Comparison with (\ref{3s})  then leads to, 
\eq{\label{311}c_1=c \alpha' \kappa^2\Lambda~,}
where we have taken (\ref{r2c}), (\ref{310}) into account. Note also that this is a check on the relative coefficient of the last two terms in (\ref{3s}).

$\bullet$  The $\phi hh$ coupling

The coupling to the dilaton can be 
obtained from (\ref{3s}) by using (\ref{31}) and keeping the terms  $\phi hh$ in the amplitude. 
A straightforward calculation, taking (\ref{etab}),  (\ref{auxr}) and  momentum conservation into account,  shows that  
the last two terms  of the  amplitude (\ref{3s}) do not contribute to the $\phi hh$ coupling. The $\phi hh$ coupling thus 
comes entirely from the first term in (\ref{3s}) upon substituting $\xi^{(s)}_{\mu\nu}\rightarrow h_{\mu\nu}$ and $\xi\rightarrow \sqrt{D-2}~ \phi$ therein. 
This is to be compared with the coupling coming from (\ref{ac}),  
\eq{
2\kappa c_1\sqrt{D-2} ~\Big( (k_3\cdot  
h_1\cdot h_{2}  \cdot k_3)~\phi_3
+\mathrm{cyclic}\Big)
~,}
and is consistent with (\ref{311}) as expected. Note that it also serves as a check of the relative coefficient between the first term in (\ref{3s}) and the last two.

$\bullet$ The  $hbb$ coupling 

The one-loop correction to the $b$-field kinetic term can be similarly derived from the $hbb$ coupling in (\ref{2b}), i.e. substituting $\xi_{\mu\nu}^{(s)}\rightarrow h_{\mu\nu}$, 
$\xi_{\mu\nu}^{(a)}\rightarrow b_{\mu\nu}$ therein. Moreover it can be seen that the trace term (proportional to $\xi_3$) and its cyclic permutations do not contribute to the coupling. 
To compare this to  (\ref{ac}), we expand $H_{\mu\nu\rho}=3\partial_{[\mu}b_{\nu\rho]}$, $G_{\mu\nu}=\eta_{\mu\nu}+2\kappa h_{\mu\nu}$  therein and keep the term $hbb$.  
After passing to momentum space, this gives,
\eq{
2\kappa ~\!c_3~\Big( 2
k_1\cdot 
h_{2}\cdot b_{3}\cdot b_{1} \cdot k_3
+2
k_1\cdot 
b_{2}\cdot h_{3}\cdot b_{1} \cdot k_2
+2
k_3\cdot 
b_{2}\cdot b_{3}\cdot h_{1} \cdot k_2
+(k_2\cdot h_{1}\cdot k_3) ~( b_{2}\cdot b_{3}) 
+\mathrm{cyclic}\Big)
~.}
Comparison with (\ref{2b}) then leads to, 
\eq{\label{315}
c_3=3c \alpha' \kappa^2\Lambda~.}
where we have taken (\ref{r2c}), (\ref{310}) into account. Note also that this is a check on the relative coefficients of the first four terms in (\ref{2b}).

$\bullet$ The $\phi bb$ coupling

Substituting $\xi_{\mu\nu}^{(s)}\rightarrow \frac{1}{\sqrt{D-2}}~\!\phi~\!\bar{\eta}_{\mu\nu}$,  
$\xi_{\mu\nu}^{(a)}\rightarrow b_{\mu\nu}$  in (\ref{2b}) we obtain,
\eq{
-6\alpha'  c g^3\Lambda
\left( \frac{4}{\sqrt{D-2}}~\!
(k_3\cdot 
b_{2}\cdot b_{3}\cdot  k_2)~\phi_1
-\sqrt{D-2}~(k_3\cdot 
b_{2}\cdot b_{3}\cdot  k_2)~\phi_1
+\mathrm{cyclic}\right)
~.}
where the first term above comes from the first four terms in (\ref{2b}) while the second term above comes from the last term in (\ref{2b}). Note that 
all $\bar{k}$-dependent terms drop out of the final result, as they should. 
The comparison with (\ref{ac}) simply provides a consistency check of (\ref{315}), with no additional information. Moreover 
it provides a check of the relative coefficient between the first four and the last term in (\ref{2b}).

$\bullet$ The $h\phi\phi$ coupling 

The $h\phi\phi$ coupling comes entirely from the last two terms in (\ref{3s}): substituting 
$\xi_{\mu\nu}^{(s)}\rightarrow \frac{1}{\sqrt{D-2}}~\!\phi~\!\bar{\eta}_{\mu\nu}$ 
for two out of the three polarizations and  $\xi^{(s)}_{\mu\nu}\rightarrow h_{\mu\nu}$ 
for the third one, we obtain,
\eq{
-2\alpha'  c g^3\Lambda
\left( ~\!\phi_2~\phi_3 ~\!
(k_3\cdot 
h_{1}\cdot   k_2)
+\mathrm{cyclic}\right)
~.}
Comparison with (\ref{ac}) then leads to,
\eq{c_2=c \alpha' \kappa^2\Lambda~.}
Assembling all previous results and rescaling: $\kappa^2\Lambda\rightarrow \Lambda$, $\kappa\phi\rightarrow \phi$, $\kappa b\rightarrow b$, 
the two-derivative  effective action  to one loop is given by, 
\eq{\spl{\label{sefffinresc}
\kappa^2\seff=  
\int\text{d}^Dx\sqrt{G}&\Big\{
(1+\gamma ~\!e^{\sqrt{D-2}~\!\!\phi})\big(
 \tfrac{1}{2}R-\tfrac12 (\partial\phi)^2\big)\\
-\tfrac16 &(1+3\gamma  ~\!e^{\sqrt{D-2}~\!\!\phi}) H^2e^{-\frac{4}{\sqrt{D-2}}~\!\!\phi}
+
e^{\frac{D}{\sqrt{D-2}}~\!\!\phi}\Lambda  
\Big\}
~,}}
where $\gamma$ is given by (\ref{cdef1}). Extrapolating to arbitrary dimension $D$, we must include the tree-level cosmological constant, which vanishes in the critical dimension $D=26$. We thus arrive at the 
effective action given in \eqref{sefffinresc1}.

\FloatBarrier

\section{Four-point amplitude, quadratic momentum}\label{sec:4p2k}

We can now start the computation of the one-loop four-point amplitude with terms quadratic in momentum. Contrary to the $N$-point amplitudes with $N\leq3$, momentum conservation 
and the on-shell condition no longer imply $k_i\cdot k_j=0$. Expanding the exponential in the four-point correlator (\ref{correlt}) as follows,
\begin{equation}
e^{-k_i\cdot k_j G_{ij}}= 1-k_i\cdot k_j G_{ij}+\dots~,
\end{equation}
we will have two terms to compute: the term coming from the 1 in the expansion of the exponential will have to be multiplied  by terms bilinear in $u_i,u_{\bar{i}}$ (i.e.~terms quadratic in momenta), while the term coming from the $k_i\cdot k_j G_{ij}$ in the expansion can only be multiplied  by terms containing $w_{ij}$'s but no $u_i$'s (i.e.~terms without additional  powers of momenta). 
\begin{figure}[h!]
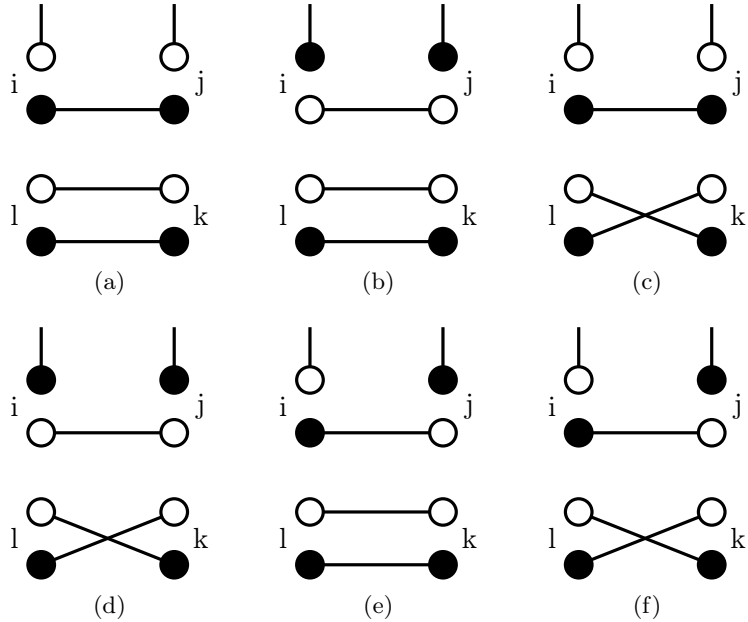

\begin{center}
\begin{tabular}{ccc}
\subfloat[]{
\tikzsetnextfilename{s1234_1p}
\input{s1234_1.tex}} 
&
\subfloat[]{
\tikzsetnextfilename{s1234_2p}
\input{s1234_2.tex}}
&
\subfloat[]{
\tikzsetnextfilename{s1234_3p}
\input{s1234_3.tex}}
\\
\subfloat[]{
\tikzsetnextfilename{s1234_4p}
\input{s1234_4.tex}}
&
\subfloat[]{
\tikzsetnextfilename{s1234_5p}
\input{s1234_5.tex}}
&
\subfloat[]{
\tikzsetnextfilename{s1234_6p}
\input{s1234_6.tex}}
\end{tabular}
\end{center}
\caption{The six admissible graphs with paired nodes corresponding to $S_{1,2,3,4}$. It is understood that the positions $i,j,k,l\in\{1,2,3,4\}$ are all different from each other.} \label{six}
\end{figure}
\FloatBarrier

\subsection{Determination of the different graphs}


{\it Terms bilinear in $u_i,u_{\bar{i}}$}

In this case, as we mentioned  earlier, there are no two-point correlators with explicit powers of momenta. 
Moreover, graphs contributing to $S_{1,2,3}$ have already been determined in sections \ref{sec:npoint}, \ref{sec:3p2k}.  
Therefore, we only need to determine the terms in $S_{1,2,3,4}$ with two  $u_i$'s. To that end, we use open vertices as we did in section \ref{sec:3p2k} to obtain the graphs of figure \ref{fig8}.

\begin{figure}[h!]
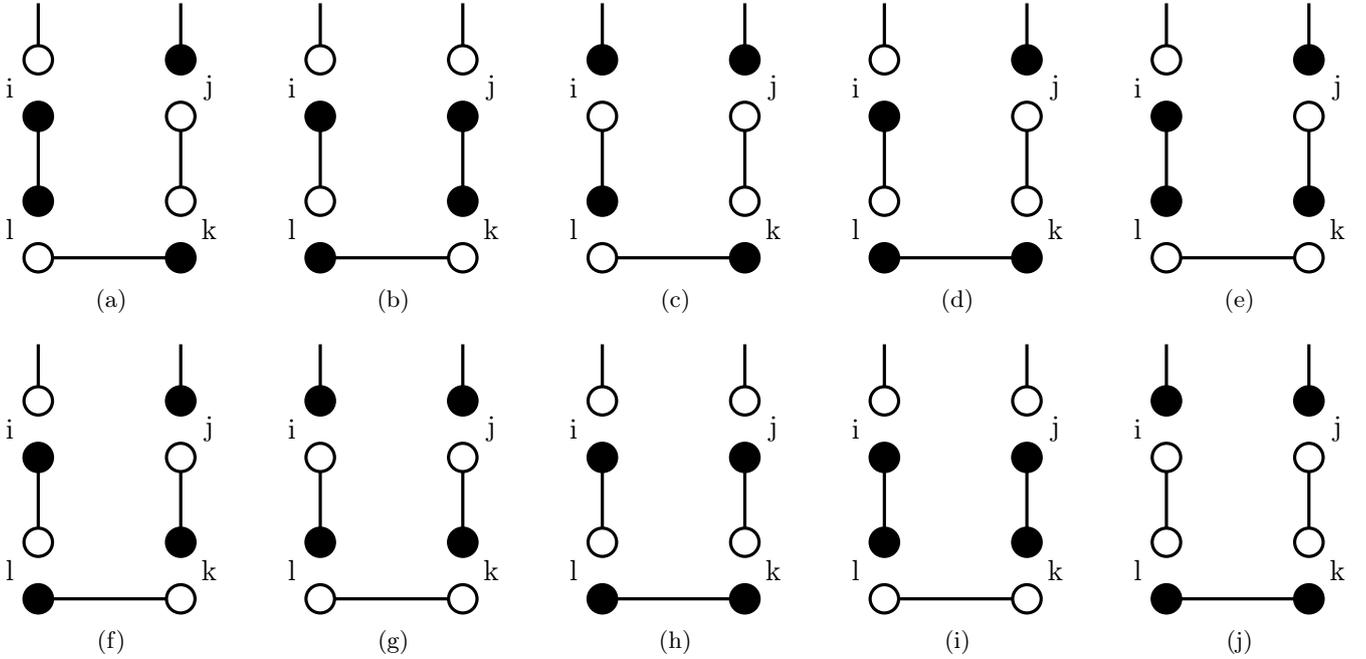

\begin{center}
\begin{tabular}{ccccc}
\subfloat[]{
\tikzsetnextfilename{s1234_ap}
\input{s1234_a.tex}} 
&
\subfloat[]{
\tikzsetnextfilename{s1234_bp}
\input{s1234_b.tex}}
&
\subfloat[]{
\tikzsetnextfilename{s1234_cp}
\input{s1234_c.tex}}
&
\subfloat[]{
\tikzsetnextfilename{s1234_dp}
\input{s1234_d.tex}}
&
\subfloat[]{
\tikzsetnextfilename{s1234_ep}
\input{s1234_e.tex}}
\\
\subfloat[]{
\tikzsetnextfilename{s1234_fp}
\input{s1234_f.tex}}
&
\subfloat[]{
\tikzsetnextfilename{s1234_gp}
\input{s1234_g.tex}}
&
\subfloat[]{
\tikzsetnextfilename{s1234_hp}
\input{s1234_h.tex}}
&
\subfloat[]{
\tikzsetnextfilename{s1234_ip}
\input{s1234_i.tex}}
&
\subfloat[]{
\tikzsetnextfilename{s1234_jp}
\input{s1234_j.tex}}

\end{tabular}
\end{center}
\caption{The ten admissible graphs with cyclic nodes corresponding to $S_{1,2,3,4}$.} \label{dix}
\end{figure}

\begin{figure}[h!]
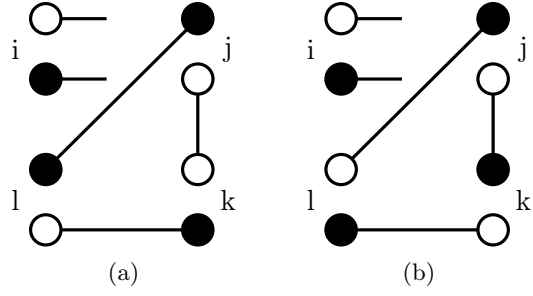

\begin{center}
\begin{tabular}{cc}
\subfloat[]{
\tikzsetnextfilename{s1234_alphap}
\input{s1234_alpha.tex}} 
&
\subfloat[]{
\tikzsetnextfilename{s1234_deltap}
\input{s1234_delta.tex}}
\end{tabular}
\end{center}
\caption{The two admissible graphs with open vertices from the same node corresponding to $S_{1,2,3,4}$.}
\label{ouvert}
\end{figure}

Performing the same operation on the graphs of the first row in figure \ref{fig7}, we obtain the six graphs of figure \ref{six}. For graphs \ref{six}.a to \ref{six}.d, we will have to sum over all pairs $i<j \in [1,4]$ and $k<l \notin \lbrace i,j \rbrace$. This gives 24 terms. 
For graphs \ref{six}.e and \ref{six}.f, we will have to sum over all $i\in [1,4], j\neq i \in [1,4]$ and $k<l \notin \lbrace i,j \rbrace$. This gives again 24 terms. 
The graphs in the second row of figure \ref{fig7} will give the ten graphs of figure \ref{dix}.
For graphs \ref{dix}.a to \ref{dix}.f, we will have to sum over all permutations. This gives $24\times 6=144$ terms. 
For graphs \ref{dix}.g to \ref{dix}.j, we will have to sum over all pairs $i<j \in [1,4]$ and $k\neq l \notin \lbrace i,j \rbrace$. This gives $12 \times 4=48$ terms.
We can also obtain graphs with two open vertices from the same node. There are two different graphs of this type depicted in fig.~\ref{ouvert}. 
For graph \ref{ouvert}.a, we will have to sum over all permutations. This gives 24 terms. 
For graph \ref{ouvert}.b, we will have to sum over all cyclic permutations and exchanges of $l$ and $k$. This gives $8$ more terms. 
We thus obtain a total of $272$ terms. We have also checked the total number of terms independently, using mathematica to expand the exponential in the four-point correlator and keep the terms bilinear in $u$ and $\xi$.  

{\it Terms proportional to $k_i\cdot k_j$}

Now let us concentrate on the terms proportional to $k_i\cdot k_j$.  
We need to determine the terms contributing to $S_{1,2}$, $S_{1,2,3}$ and $S_{1,2,3,4}$. 
The corresponding graphs were determined earlier and are given in figs.~\ref{fig6}, \ref{fig7}.

%
%
\subsection{Computation of the different graph contributions}

{\it Terms bilinear in $u_i,u_{\bar{i}}$}

All terms will be computed in a similar way using the infinite sums of section \ref{sec:3p2k}; no new sums appear at four points. 
Let us do the complete computation of the first term in fig.~\ref{six}a. 
We have to compute $u_iu_jw_{\bar{i}\bar{j}}w_{kl}w_{\bar{k}\bar{l}}$ with $i,j,k,l$ all different in $\lbrace 1,2,3,4 \rbrace$. There are nine terms coming from,
\begin{align}
u_ i u_j w_{\bar{i}\bar{j}}w_{kl}w_{\bar{k}\bar{l}}  =& -\xi_i\cdot \left( k_j\partial_iG_{ij}+k_k\partial_iG{ik}+k_l\partial_iG_{il}\right) \nonumber \\
& \times \xi_j \cdot \left( k_i \partial_j G_{ji}+k_k\partial_jG{jk}+k_l\partial_jG_{jl}\right) \nonumber \\
& \times \left( \bar{\xi}_i \cdot \bar{\xi}_j \right)\left(\bar{\partial}_i \bar{\partial}_j G_{ij}\right)\left(\xi_k \cdot \xi_l \right) \left( \partial_k \partial_l G_{kl} \right) \left(\bar{\xi}_k \cdot \bar{\xi} _l\right)\left(\bar{\partial}_k \bar{\partial}_l G_{kl}\right)~.
\end{align}
The corresponding kinematic term is proportional to  $(\xi_i\cdot k_j)(\xi_j\cdot k_i)\left( \bar{\xi}_i \cdot \bar{\xi}_j \right)\left(\xi_k \cdot \xi_l \right)\left(\bar{\xi}_k \cdot \bar{\xi} _l\right)$.  
Moreover we calculate, 
\begin{align}
&(\partial_iG_{ij})(\partial_jG_{ji})\left(\bar{\partial}_i \bar{\partial}_j G_{ij}\right)\left( \partial_k \partial_l G_{kl} \right)\left(\bar{\partial}_k \bar{\partial}_l G_{kl}\right)= \nonumber \\
&-\left(\frac{\alpha'}{4\pi^2}\right)^2\left(\frac{\alpha'}{4\pi\tau_2}\right)^3\tau_2^{5/2}\sum^{\prime}_{
\substack{(m_1,n_1)\\ (m_2,n_2)\\(m_3,n_3)\\ (m_4,n_4)\\(m_5,n_5)}}\frac{(m_1-n_1\bar{\tau})(m_2-n_2\bar{\tau})(m_3-n_3\tau)^2(m_4-n_4\bar{\tau})^2(m_5-n_5\tau)^2}{|m_1-n_1\tau|^2|m_2-n_2\tau|^2|m_3-n_3\tau|^2|m_4-n_4\tau|^2|m_5-n_5\tau|^2} \nonumber \\
& \times \psi_{m_1,n_1}(z_{ij})\psi_{m_2,n_2}(z_{ji})\psi_{m_3,n_3}(z_{ij})\psi_{m_4,n_4}(z_{kl})\psi_{m_5,n_5}(z_{kl})
~,
\end{align}
where as usual the prime over the sum means that we exclude the $(m_i,n_i)=(0,0)$ terms. 
After integration over the vertex positions $z_j,z_k,z_l$, we obtain, using (\ref{25}):
\begin{align}
&-\left(\frac{\alpha'}{8\pi}\right)^5\frac{1}{\pi^2}\sum^{\prime}_{\substack{(m_1,n_1) \\ (m_2,n_2) \\ (m_1,n_1) \neq (m_2,n_2)}}\frac{(m_1-n_1\bar{\tau})(m_2-n_2\bar{\tau})((m_2-m_1)-(n_2-n_1)\tau)^2}{|m_1-n_1\tau|^2|m_2-n_2\tau|^2|(m_2-m_1)-(n_2-n_1)\tau|^2}\sum^{\prime}_{(m_4,n_4)}1 \nonumber \\
&=-\left(\frac{\alpha'}{8\pi}\right)^5\frac{1}{\pi^2}\frac{2}{\tau_2}E_1(\tau)E_0(\tau) 
=\left(\frac{\alpha'}{8\pi}\right)^5\frac{2}{\pi^2\tau_2}E_1(\tau)~.
\end{align}
We compute the other terms in a similar way. The contribution from the graph \ref{six}.a reads,
\begin{align}
\left(\frac{\alpha'}{8\pi}\right)^5\frac{E_1(\tau)}{\pi^2\tau_2}\Big[-3(k_l\cdot\xi_i\cdot\xi_j^T\cdot k_k+k_k\cdot\xi_i\cdot\xi_j^T\cdot k_l) 
-k_k\cdot\xi_i\cdot\xi_j^T\cdot k_k-k_l\cdot\xi_i\cdot\xi_j^T\cdot k_l\Big](\xi_k\cdot\xi_l^T)~,
\end{align}
where we have used momentum conservation and recombined $\xi_i^{\mu}\bar{\xi}_i^{\nu}\rightarrow\xi_i^{\mu\nu}$.

To sum over the indices, we decompose the polarizations into their symmetric and anti-symmetric parts. 
By adding \ref{six}.a with \ref{six}.b and \ref{six}.c with \ref{six}.d we thus obtain terms of the form,
\begin{align}
&k_k\cdot(\xi_i\cdot\xi_j^T+\xi_i^T\cdot\xi_j)\cdot k_l=2k_k\cdot(\xi_i^{(s)}\xi_j^{(s)}-\xi_i^{(a)}\xi_j^{(a)})\cdot k_l
~.
\end{align}
Indeed, we can obtain the contribution of \ref{six}.b from the one of \ref{six}.a by exchanging $\xi$ and $\bar{\xi}$. This is equivalent to exchanging: $\xi_{\mu\nu} \leftrightarrow \xi_{\nu\mu}$. 

Summing over the indices in \ref{six}.e and \ref{six}.f, we obtain terms of the form,
\begin{align}
&k_k\cdot(\xi_i\cdot\xi_j+\xi_j\cdot\xi_i)\cdot k_k=k_k\cdot(\xi_i\cdot\xi_j+\xi_i^T\cdot\xi_j^T)\cdot k_k  
=2k_k\cdot(\xi_i^{(s)}\cdot\xi_j^{(s)}+\xi_i^{(a)}\cdot\xi_j^{(a)})\cdot k_k~.
\end{align}
Moreover  we have,
\begin{align}
\xi_k\cdot\xi_l&=\xi_k^{(s)}\cdot\xi_l^{(s)}+\xi_k^{(a)}\cdot \xi_l^{(a)} \nonumber \\
 \xi_k\cdot\xi_l^T&=\xi_k^{(s)}\cdot\xi_l^{(s)}-\xi_k^{(a)}\cdot \xi_l^{(a)} ~.
\end{align}
Finally, all the contributions from figure \ref{six} give the following term, 
%
\begin{align}
\left(\frac{\alpha'}{8\pi}\right)^5\frac{E_1(\tau)}{\pi^2\tau_2}&\left[4k_3\cdot \xi_1^{(s)}\cdot \xi_2^{(s)}\cdot k_3+12k_3\cdot \xi_1^{(a)}\cdot \xi_2^{(a)}\cdot k_3 \right.+12k_3\cdot \xi_1^{(a)}\cdot \xi_2^{(a)}\cdot k_4 \nonumber \\
& \left. +4k_4\cdot \xi_1^{(s)}\cdot \xi_2^{(s)}\cdot k_4
+12k_4\cdot \xi_1^{(a)}\cdot \xi_2^{(a)}\cdot k_4 +12k_4\cdot \xi_1^{(a)}\cdot \xi_2^{(a)}\cdot k_3 \right](\xi_3^{(s)}\cdot \xi_4^{(s)}) \nonumber \\
&+\left[  12k_3\cdot \xi_1^{(s)}\cdot \xi_2^{(s)}\cdot k_4+12k_4\cdot \xi_1^{(s)}\cdot \xi_2^{(s)}\cdot k_3\right](\xi_3^{(a)}\cdot \xi_4^{(a)}) 
+\text{other pairs}~,
\label{finalsix}
\end{align}
where by ``other pairs'' we mean that there are six terms in total: one for each of the following pairs:  (12/34), (13/24), (14/23), (23/14), (24/13) and (34/12).
%
%
\vfill\break

The contribution from the graphs \ref{dix}.b is computed in a very similar way. 
We obtain the following, 
%
\begin{align}
\left(\frac{\alpha'}{8\pi}\right)^5\frac{E_1(\tau)}{\pi^2\tau_2}&\Big[k_3 \cdot \left( 4\xi_1^{(s)} \cdot \xi_4^{(s)} \cdot \xi_3^{(s)} \cdot \xi_2^{(s)} -12 \xi_1^{(a)} \cdot \xi_4^{(a)} \cdot \xi_3^{(a)} \cdot \xi_2^{(a)} +12\xi_1^{(a)} \cdot \xi_4^{(a)} \cdot \xi_3^{(s)} \cdot \xi_2^{(s)}\right.  \nonumber \\
&\hspace{1cm}\left.+12\xi_1^{(a)} \cdot \xi_4^{(s)} \cdot \xi_3^{(a)} \cdot \xi_2^{(s)}  -12\xi_1^{(a)} \cdot \xi_4^{(s)} \cdot \xi_3^{(s)} \cdot \xi_2^{(a)} -12\xi_1^{(s)} \cdot \xi_4^{(a)} \cdot \xi_3^{(a)} \cdot \xi_2^{(s)}\right.  \nonumber \\
&\hspace{1cm}\left.+12\xi_1^{(s)} \cdot \xi_4^{(a)} \cdot \xi_3^{(s)} \cdot \xi_2^{(a)}+12\xi_1^{(s)} \cdot \xi_4^{(s)} \cdot \xi_3^{(a)} \cdot \xi_2^{(a)}\right)\cdot k_3 \nonumber \\
&+k_4 \cdot \left( -20\xi_1^{(s)} \cdot \xi_4^{(s)} \cdot \xi_3^{(s)} \cdot \xi_2^{(s)} +12 \xi_1^{(a)} \cdot \xi_4^{(a)} \cdot \xi_3^{(a)} \cdot \xi_2^{(a)} +12\xi_1^{(a)} \cdot \xi_4^{(a)} \cdot \xi_3^{(s)} \cdot \xi_2^{(s)}\right.  \nonumber \\
&\hspace{1cm}\left. +12\xi_1^{(a)} \cdot \xi_4^{(s)} \cdot \xi_3^{(a)} \cdot \xi_2^{(s)}   -36\xi_1^{(a)} \cdot \xi_4^{(s)} \cdot \xi_3^{(s)} \cdot \xi_2^{(a)} +12\xi_1^{(s)} \cdot \xi_4^{(a)} \cdot \xi_3^{(a)} \cdot \xi_2^{(s)}\right.  \nonumber \\
&\hspace{1cm}\left.+12\xi_1^{(s)} \cdot \xi_4^{(a)} \cdot \xi_3^{(s)} \cdot \xi_2^{(a)}+12\xi_1^{(s)} \cdot \xi_4^{(s)} \cdot \xi_3^{(a)} \cdot \xi_2^{(a)}\right)\cdot k_4 \nonumber \\
&+k_3 \cdot \left( 2\xi_1^{(s)} \cdot \xi_4^{(s)} \cdot \xi_3^{(s)} \cdot \xi_2^{(s)} +2 \xi_1^{(a)} \cdot \xi_4^{(a)} \cdot \xi_3^{(a)} \cdot \xi_2^{(a)} +-2\xi_1^{(a)} \cdot \xi_4^{(a)} \cdot \xi_3^{(s)} \cdot \xi_2^{(s)}\right.  \nonumber \\
&\hspace{1cm}\left.+22\xi_1^{(a)} \cdot \xi_4^{(s)} \cdot \xi_3^{(a)} \cdot \xi_2^{(s)}  -22\xi_1^{(a)} \cdot \xi_4^{(s)} \cdot \xi_3^{(s)} \cdot \xi_2^{(a)} -22\xi_1^{(s)} \cdot \xi_4^{(a)} \cdot \xi_3^{(a)} \cdot \xi_2^{(s)}\right.  \nonumber \\
&\hspace{1cm}\left.+22\xi_1^{(s)} \cdot \xi_4^{(a)} \cdot \xi_3^{(s)} \cdot \xi_2^{(a)}-2\xi_1^{(s)} \cdot \xi_4^{(s)} \cdot \xi_3^{(a)} \cdot \xi_2^{(a)}\right)\cdot k_4 \nonumber \\
&+k_4 \cdot \left( -8\xi_1^{(s)} \cdot \xi_4^{(s)} \cdot \xi_3^{(s)} \cdot \xi_2^{(s)} +24 \xi_1^{(s)} \cdot \xi_4^{(s)} \cdot \xi_3^{(a)} \cdot \xi_2^{(a)} -2\xi_1^{(s)} \cdot \xi_4^{(a)} \cdot \xi_3^{(a)} \cdot \xi_2^{(s)} \right.  \nonumber \\
&\hspace{1cm}\left.  -24\xi_1^{(a)} \cdot \xi_4^{(s)} \cdot \xi_3^{(s)} \cdot \xi_2^{(a)} +24\xi_1^{(a)} \cdot \xi_4^{(a)} \cdot \xi_3^{(s)} \cdot \xi_2^{(s)}\right)\cdot k_3  
 + 1 \leftrightarrow 2 \Big] +  \text{other pairs}~.
\label{finaldix}
\end{align}
Similarly, we obtain the contribution from all the graphs in figure \ref{ouvert}, 
%
\begin{align}
\left(\frac{\alpha'}{8\pi}\right)^5\frac{E_1(\tau)}{\pi^2\tau_2}&\left[8k_3\cdot \xi_1 \cdot k_3+8k_4\cdot \xi_1 \cdot k_4+4k_4\cdot \xi_1 \cdot k_3+4k_3\cdot \xi_1 \cdot k_4\right] \nonumber \\
&~~~~~~~~~~~~~~~~~~~~~~~~~~~~~~~~~~~~~~~~~~~~~~~
\times \left( \xi_2^{(s)}\cdot \xi_4^{(s)}\cdot \xi_3^{(s)} + \xi_2^{(s)}\cdot \xi_4^{(a)}\cdot \xi_3^{(a)} \right) \nonumber \\
&+\left[-16k_3\cdot \xi_1 \cdot k_3-16k_4\cdot \xi_1 \cdot k_4-8k_4\cdot \xi_1 \cdot k_3-8k_3\cdot \xi_1 \cdot k_4\right]  \nonumber \\
&~~~~~~~~~~~~~~~~~~~~~~~~~~~~~~~~~~~~~~~~~~~~~~~
\times \left(\xi_2^{(a)}\cdot \xi_4^{(s)}\cdot \xi_3^{(a)}+\xi_2^{(a)}\cdot \xi_4^{(a)}\cdot \xi_3^{(s)}\right) + \text{cyclic}~.
\end{align}
This concludes the computation of all the terms in $S_{1,2,3,4}$ quadratic in momentum.

It is also necessary to redo the computation of $S_{1,2,3}$ and $S_{1,2}$ because we now have four possible momenta $k_i$ and one more position integration than for the three-point amplitude. We will refrain from giving explicit details of the computations, which are very similar to the previous ones. The results are summarized in the following equations. 
\begin{align}
S_{1,2,3}=&\frac{1}{\pi^2}\left(\frac{\alpha'}{8\pi}\right)^4E_1(\tau)\Big[  k_3\cdot (-4\xi_2^{(s)}\cdot\xi_3^{(s)}\cdot\xi_1^{(s)}+12\xi_2^{(s)}\cdot\xi_3^{(a)}\cdot\xi_1^{(a)}+12\xi_2^{(a)}\cdot\xi_3^{(a)}\cdot\xi_1^{(s)})\cdot k_3 \nonumber \\
&+k_3\cdot (-6\xi_2^{(s)}\cdot\xi_3^{(s)}\cdot\xi_1^{(s)}-6\xi_2^{(a)}\cdot\xi_3^{(s)}\cdot\xi_1^{(a)}+6\xi_2^{(s)}\cdot\xi_3^{(a)}\cdot\xi_1^{(a)}+6\xi_2^{(a)}\cdot\xi_3^{(a)}\cdot\xi_1^{(s)})\cdot k_4 \nonumber \\
&+k_4\cdot (6\xi_2^{(s)}\cdot\xi_3^{(s)}\cdot\xi_1^{(s)}-18\xi_2^{(a)}\cdot\xi_3^{(s)}\cdot\xi_1^{(a)}+6\xi_2^{(s)}\cdot\xi_3^{(a)}\cdot\xi_1^{(a)}+6\xi_2^{(a)}\cdot\xi_3^{(a)}\cdot\xi_1^{(s)})\cdot k_3 \nonumber \\
&+k_4\cdot (12\xi_2^{(s)}\cdot\xi_3^{(s)}\cdot\xi_1^{(s)}-12\xi_2^{(a)}\cdot\xi_3^{(s)}\cdot\xi_1^{(a)})\cdot k_4 \nonumber \\
&+(2k_3\cdot\xi_1\cdot k_3 + k_3\cdot\xi_1\cdot k_4+ k_4\cdot\xi_1\cdot k_3 +4k_4\cdot\xi_1\cdot k_4)(\xi_2^{(s)}\cdot\xi_3^{(s)}) \nonumber \\
&\left. +(-6k_3\cdot\xi_1\cdot k_3 -3 k_3\cdot\xi_1\cdot k_4-3k_4\cdot\xi_1\cdot k_3)(\xi_2^{(a)}\cdot\xi_3^{(a)}) + \text{cyclic} \right]~.
\label{s123}
\end{align}
\FloatBarrier
Similarly, we obtain the new expression for $S_{1,2}$ \eqref{s12}:
\begin{align}
S_{1,2}=&\frac{1}{\pi^2}\left(\frac{\alpha'}{8\pi}\right)^4E_1(\tau)\tau_2\left[ k_3\cdot( -2\xi_1^{(s)}\cdot \xi_2^{(s)}-6\xi_1^{(a)}\cdot \xi_2^{(a)})\cdot k_3+ k_3\cdot( 2\xi_1^{(s)}\cdot \xi_2^{(s)}-6\xi_1^{(a)}\cdot \xi_2^{(a)})\cdot k_4 \right. \nonumber \\
&+ k_4\cdot( 2\xi_1^{(s)}\cdot \xi_2^{(s)}-6\xi_1^{(a)}\cdot \xi_2^{(a)})\cdot k_3 + \left. k_4\cdot( -2\xi_1^{(s)}\cdot \xi_2^{(s)}-6\xi_1^{(a)}\cdot \xi_2^{(a)})\cdot k_4 \right]
\label{s12}
\end{align}
{\it Terms proportional to $k_i\cdot k_j$}

We have to compute the terms corresponding to  the graphs in figures \ref{fig3}, \ref{fig4} and \ref{fig7}. 
The polarization terms and the coefficients are computed as before. 
We obtain the following results.
\begin{align} 
(\prod_{i<j}k_i\cdot k_j G_{ij})S_{1,2,3,4}=\frac{1}{\pi^2\tau_2}\left(\frac{\alpha'}{8\pi}\right)^5E_1(\tau)\Big[  & \left\lbrace 12(k_1\cdot k_2)(\xi_1^{(s)}\xi_4^{(a)}\xi_3^{(s)}\xi_2^{(a)}-\xi_1^{(s)}\xi_4^{(a)}\xi_3^{(a)}\xi_2^{(s)}\right.  \nonumber \\
&-\xi_1^{(a)}\xi_4^{(s)}\xi_3^{(s)}\xi_2^{(a)}+\xi_1^{(a)}\xi_4^{(a)}\xi_3^{(a)}\xi_2^{(s)}) \nonumber \\
&+12(k_1\cdot k_4)(-\xi_1^{(s)}\xi_4^{(s)}\xi_3^{(a)}\xi_2^{(a)}+\xi_1^{(s)}\xi_4^{(a)}\xi_3^{(s)}\xi_2^{a)}  \nonumber \\
&\left. +\xi_1^{(a)}\xi_4^{(s)}\xi_3^{(a)}\xi_2^{(s)}-\xi_1^{(a)}\xi_4^{(a)}\xi_3^{(s)}\xi_2^{(s)}) + 3\leftrightarrow 4 + 2 \leftrightarrow 3 \right\rbrace \nonumber \\
&+\left\lbrace 2(k_1 \cdot k_2)(4(\xi_1^{(s)}\xi_2^{(s)})(\xi_3^{(s)}\xi_4^{(s)})+(\xi_1^{(s)}\xi_2^{(s)})(\xi_3^{(a)}\xi_4^{(a)}) \right. \nonumber \\
&\left. \left. +(\xi_1^{(a)}\xi_2^{(a)})(\xi_3^{(s)}\xi_4^{(s)})+2(\xi_1^{(a)}\xi_2^{(a)})(\xi_3^{(a)}\xi_4^{(a)})+ 2\leftrightarrow 3 + 2\leftrightarrow 4 \right\rbrace \right]
\end{align}
\begin{align}\label{a41}
(\prod_{i<j}k_i\cdot k_j G_{ij})S_{1,2,3}=\frac{12}{\pi^2}\left(\frac{\alpha'}{8\pi}\right)^4E_1(\tau)\Big[&(k_1\cdot k_4)(-\xi_1^{(s)}\cdot\xi_3^{(a)}\cdot \xi_2^{(a)}+\xi_1^{(a)}\cdot\xi_3^{(a)}\cdot \xi_2^{(s)}) \nonumber \\
&\left. +(k_1\cdot k_2)(-\xi_1^{(a)}\cdot\xi_3^{(s)}\cdot \xi_2^{(a)}+\xi_1^{(a)}\cdot\xi_3^{(a)}\cdot \xi_2^{(s)})\right]
\end{align}
\begin{align}\label{a42}
(\prod_{i<j}k_i\cdot k_j G_{ij})S_{1,2}=&\frac{\tau_2}{\pi^2}\left(\frac{\alpha'}{8\pi}\right)^3E_1(\tau)(k_1\cdot k_2)\left[-(\xi_1^{(s)}\cdot\xi_2^{(s)})-3(\xi_1^{(a)}\cdot\xi_2^{(a)})\right]
\end{align}

{\it Final result}

We can now substitute all the previous results in eqs.~\eqref{correlt},  \eqref{ampl} to arrive at the four-point amplitude 
quadratic in momenta, 

\vfill\break

{\small
\begin{align}\label{a4k2}
&\mathcal{A}_4^{k^2} =-\gamma \kappa^4
\left[ \textcolor{white}{\int}\right.  \nonumber \\
&\left\lbrace\left[4k_3\cdot \xi_1^{(s)}\cdot \xi_2^{(s)}\cdot k_3+12k_3\cdot \xi_1^{(a)}\cdot \xi_2^{(a)}\cdot k_3 \right. +12k_3\cdot \xi_1^{(a)}\cdot \xi_2^{(a)}\cdot k_4 \right.\nonumber \\
& \left. +4k_4\cdot \xi_1^{(s)}\cdot \xi_2^{(s)}\cdot k_4
+12k_4\cdot \xi_1^{(a)}\cdot \xi_2^{(a)}\cdot k_4 +12k_4\cdot \xi_1^{(a)}\cdot \xi_2^{(a)}\cdot k_3 \right](\xi_3^{(s)}\cdot \xi_4^{(s)}) \nonumber \\
&+\left[ 12k_3\cdot \xi_1^{(s)}\cdot \xi_2^{(s)}\cdot k_4+12k_4\cdot \xi_1^{(s)}\cdot \xi_2^{(s)}\cdot k_3 \right](\xi_3^{(a)}\cdot \xi_4^{(a)}) \nonumber \\
&+\left[k_3 \cdot \left( 4\xi_1^{(s)} \cdot \xi_4^{(s)} \cdot \xi_3^{(s)} \cdot \xi_2^{(s)} -12 \xi_1^{(a)} \cdot \xi_4^{(a)} \cdot \xi_3^{(a)} \cdot \xi_2^{(a)} +12\xi_1^{(a)} \cdot \xi_4^{(a)} \cdot \xi_3^{(s)} \cdot \xi_2^{(s)}\right. \right. \nonumber \\
&\hspace{1cm}\left.+12\xi_1^{(a)} \cdot \xi_4^{(s)} \cdot \xi_3^{(a)} \cdot \xi_2^{(s)}  -12\xi_1^{(a)} \cdot \xi_4^{(s)} \cdot \xi_3^{(s)} \cdot \xi_2^{(a)} -12\xi_1^{(s)} \cdot \xi_4^{(a)} \cdot \xi_3^{(a)} \cdot \xi_2^{(s)}\right.  \nonumber \\
&\hspace{1cm}\left.+12\xi_1^{(s)} \cdot \xi_4^{(a)} \cdot \xi_3^{(s)} \cdot \xi_2^{(a)}+12\xi_1^{(s)} \cdot \xi_4^{(s)} \cdot \xi_3^{(a)} \cdot \xi_2^{(a)}\right)\cdot k_3 \nonumber \\
&+k_4 \cdot \left( -20\xi_1^{(s)} \cdot \xi_4^{(s)} \cdot \xi_3^{(s)} \cdot \xi_2^{(s)} +12 \xi_1^{(a)} \cdot \xi_4^{(a)} \cdot \xi_3^{(a)} \cdot \xi_2^{(a)} +12\xi_1^{(a)} \cdot \xi_4^{(a)} \cdot \xi_3^{(s)} \cdot \xi_2^{(s)}\right.  \nonumber \\
&\hspace{1cm}\left. +12\xi_1^{(a)} \cdot \xi_4^{(s)} \cdot \xi_3^{(a)} \cdot \xi_2^{(s)}   -36\xi_1^{(a)} \cdot \xi_4^{(s)} \cdot \xi_3^{(s)} \cdot \xi_2^{(a)} +12\xi_1^{(s)} \cdot \xi_4^{(a)} \cdot \xi_3^{(a)} \cdot \xi_2^{(s)}\right.  \nonumber \\
&\hspace{1cm}\left.+12\xi_1^{(s)} \cdot \xi_4^{(a)} \cdot \xi_3^{(s)} \cdot \xi_2^{(a)}+12\xi_1^{(s)} \cdot \xi_4^{(s)} \cdot \xi_3^{(a)} \cdot \xi_2^{(a)}\right)\cdot k_4 \nonumber \\
&+k_3 \cdot \left( 2\xi_1^{(s)} \cdot \xi_4^{(s)} \cdot \xi_3^{(s)} \cdot \xi_2^{(s)} +2 \xi_1^{(a)} \cdot \xi_4^{(a)} \cdot \xi_3^{(a)} \cdot \xi_2^{(a)} -2\xi_1^{(a)} \cdot \xi_4^{(a)} \cdot \xi_3^{(s)} \cdot \xi_2^{(s)}\right.  \nonumber \\
&\hspace{1cm}\left.+22\xi_1^{(a)} \cdot \xi_4^{(s)} \cdot \xi_3^{(a)} \cdot \xi_2^{(s)}  -22\xi_1^{(a)} \cdot \xi_4^{(s)} \cdot \xi_3^{(s)} \cdot \xi_2^{(a)} -22\xi_1^{(s)} \cdot \xi_4^{(a)} \cdot \xi_3^{(a)} \cdot \xi_2^{(s)}\right.  \nonumber \\
&\hspace{1cm}\left.+22\xi_1^{(s)} \cdot \xi_4^{(a)} \cdot \xi_3^{(s)} \cdot \xi_2^{(a)}-2\xi_1^{(s)} \cdot \xi_4^{(s)} \cdot \xi_3^{(a)} \cdot \xi_2^{(a)}\right)\cdot k_4 \nonumber \\
&+k_4 \cdot \left( -8\xi_1^{(s)} \cdot \xi_4^{(s)} \cdot \xi_3^{(s)} \cdot \xi_2^{(s)} +24 \xi_1^{(s)} \cdot \xi_4^{(s)} \cdot \xi_3^{(a)} \cdot \xi_2^{(a)} -2\xi_1^{(s)} \cdot \xi_4^{(a)} \cdot \xi_3^{(a)} \cdot \xi_2^{(s)} \right.  \nonumber \\
&\hspace{1cm}\left.  -24\xi_1^{(a)} \cdot \xi_4^{(s)} \cdot \xi_3^{(s)} \cdot \xi_2^{(a)} +24\xi_1^{(a)} \cdot \xi_4^{(a)} \cdot \xi_3^{(s)} \cdot \xi_2^{(s)}\right)\cdot k_3  
 + 1 \leftrightarrow 2 \Big] +  \text{other pairs}\Big\rbrace \nonumber \\
&+\Big\lbrace\left[8k_3\cdot \xi_1 \cdot k_3+8k_4\cdot \xi_1 \cdot k_4+4k_4\cdot \xi_1 \cdot k_3+4k_3\cdot \xi_1 \cdot k_4\right] \left( \xi_2^{(s)}\cdot \xi_4^{(s)}\cdot \xi_3^{(s)} + \xi_2^{(s)}\cdot \xi_4^{(a)}\cdot \xi_3^{(a)} \right) \nonumber \\
&+\left[-16k_3\cdot \xi_1 \cdot k_3-16k_4\cdot \xi_1 \cdot k_4-8k_4\cdot \xi_1 \cdot k_3-8k_3\cdot \xi_1 \cdot k_4\right] \left.  \left(\xi_2^{(a)}\cdot \xi_4^{(s)}\cdot \xi_3^{(a)}+\xi_2^{(a)}\cdot \xi_4^{(a)}\cdot \xi_3^{(s)}\right) + \text{cyclic} \right\rbrace \nonumber \\
&-\left\lbrace \xi_4 \left[ k_3\cdot (-4\xi_2^{(s)}\cdot\xi_3^{(s)}\cdot\xi_1^{(s)}+12\xi_2^{(a)}\cdot\xi_3^{(a)}\cdot\xi_1^{(s)}-12\xi_2^{(a)}\cdot\xi_3^{(s)}\cdot\xi_1^{(a)}+12\xi_2^{(s)}\cdot\xi_3^{(a)}\cdot\xi_1^{(a)})\cdot k_3 \right. \right. \nonumber \\
&\hspace{1cm}+k_3\cdot (-12\xi_2^{(a)}\cdot\xi_3^{(s)}\cdot\xi_1^{(a)}+12\xi_2^{(a)}\cdot\xi_3^{(a)}\cdot\xi_1^{(s)})\cdot k_4 +k_4\cdot (-12\xi_2^{(a)}\cdot\xi_3^{(s)}\cdot\xi_1^{(a)}+12\xi_2^{(s)}\cdot\xi_3^{(a)}\cdot\xi_1^{(a)})\cdot k_3 \nonumber \\
& \hspace{1cm}+k_4\cdot (-4\xi_2^{(s)}\cdot\xi_3^{(s)}\cdot\xi_1^{(s)}-12\xi_2^{(a)}\cdot\xi_3^{(s)}\cdot\xi_1^{(a)})\cdot k_4 \nonumber \\
&\hspace{1cm}+(2k_3\cdot\xi_1\cdot k_3 + k_3\cdot\xi_1\cdot k_4+ k_4\cdot\xi_1\cdot k_3 +4k_4\cdot\xi_1\cdot k_4)(\xi_2^{(s)}\cdot\xi_3^{(s)}) \nonumber \\
&\hspace{1cm}\left. \left. +(-6k_3\cdot\xi_1\cdot k_3 -3 k_3\cdot\xi_1\cdot k_4-3k_4\cdot\xi_1\cdot k_3)(\xi_2^{(a)}\cdot\xi_3^{(a)}) + \text{cyclic in 123} \right] +\text{cyclic in 1234} \right\rbrace \nonumber \\
&+\left\lbrace \xi_3\xi_4 \left(k_3\cdot( -2\xi_1^{(s)}\cdot \xi_2^{(s)}-6\xi_1^{(a)}\cdot \xi_2^{(a)})\cdot k_3+ k_3\cdot( 2\xi_1^{(s)}\cdot \xi_2^{(s)}-6\xi_1^{(a)}\cdot \xi_2^{(a)})\cdot k_4 \right.\right.\nonumber \\
&\hspace{1.5cm} \left.\left. +k_4\cdot( 2\xi_1^{(s)}\cdot \xi_2^{(s)}-6\xi_1^{(a)}\cdot \xi_2^{(a)})\cdot k_3 +  k_4\cdot( -2\xi_1^{(s)}\cdot \xi_2^{(s)}-6\xi_1^{(a)}\cdot \xi_2^{(a)})\cdot k_4 \right) +\text{other pairs} \right\rbrace  \nonumber \\
&-\left\lbrace 12(k_1\cdot k_2)(\xi_1^{(s)}\xi_4^{(a)}\xi_3^{(s)}\xi_2^{(a)}-\xi_1^{(s)}\xi_4^{(a)}\xi_3^{(a)}\xi_2^{(s)}-\xi_1^{(a)}\xi_4^{(s)}\xi_3^{(s)}\xi_2^{(a)}+\xi_1^{(a)}\xi_4^{(s)}\xi_3^{(a)}\xi_2^{(s)}) \right. \nonumber \\
&\hspace{0.5cm}+12(k_1\cdot k_4)(-\xi_1^{(s)}\xi_4^{(s)}\xi_3^{(a)}\xi_2^{(a)}+\xi_1^{(s)}\xi_4^{(a)}\xi_3^{(s)}\xi_2^{a)}  \left. +\xi_1^{(a)}\xi_4^{(s)}\xi_3^{(a)}\xi_2^{(s)}-\xi_1^{(a)}\xi_4^{(a)}\xi_3^{(s)}\xi_2^{(s)}) + 3\leftrightarrow 4 + 2 \leftrightarrow 3 \right\rbrace \nonumber \\
&-\left\lbrace 2(k_1 \cdot k_2)(4(\xi_1^{(s)}\xi_2^{(s)})(\xi_3^{(s)}\xi_4^{(s)})+(\xi_1^{(s)}\xi_2^{(s)})(\xi_3^{(a)}\xi_4^{(a)})  +(\xi_1^{(a)}\xi_2^{(a)})(\xi_3^{(s)}\xi_4^{(s)})+2(\xi_1^{(a)}\xi_2^{(a)})(\xi_3^{(a)}\xi_4^{(a)}))+ 2\leftrightarrow 3,4  \right\rbrace \nonumber \\
&+12\left\lbrace\xi_4(k_1\cdot k_4)(-\xi_1^{(s)}\cdot\xi_3^{(a)}\cdot \xi_2^{(a)}+\xi_1^{(a)}\cdot\xi_3^{(a)}\cdot \xi_2^{(s)}) 
+\xi_4(k_1\cdot k_2)(-\xi_1^{(a)}\cdot\xi_3^{(s)}\cdot \xi_2^{(a)}+\xi_1^{(a)}\cdot\xi_3^{(a)}\cdot \xi_2^{(s)})+ \text{cyclic}\right\rbrace \nonumber \\
& +\left\lbrace\xi_3\xi_4(k_1\cdot k_2)\left[(\xi_1^{(s)}\cdot\xi_2^{(s)})+3(\xi_1^{(a)}\cdot\xi_2^{(a)})\right]+\text{other pairs}\right\rbrace 
\left. \textcolor{white}{\int}
\hskip-.3cm \right]~,
\end{align}
}
where we took \eqref{ldef}, \eqref{r2c} into account, and we have rescaled $g\rightarrow -g\alpha'/8\pi$ as usual; we have also used \eqref{310} to substitute $\kappa$ for $g$.

\subsection{Comparison with the effective action}\label{sec:comp}

In this subsection we will compare some of the terms in $\mathcal{A}_4^{k^2}$, cf.~\eqref{a4k2}, with the effective action \eqref{sefffinresc1}. 
We do not attempt a complete comparison, which 
would be rather involved and requires systematically taking into account all 1-particle reducible graphs. In the following we will only examine the terms in the amplitude coming from 
\eqref{a41}, which should be compared with the terms in the effective action linear in the dilaton,  and \eqref{a42} which should be compared with the terms quadratic in the dilaton.

{\it Terms of the form $(k_i\cdot k_j)\phi h^3$}

Expanding the linear coupling of the Einstein term to the dilaton around flat space to cubic order in $h$ and passing to momentum space, we obtain,
\eq{\spl{
\sqrt{G} R\phi\rightarrow4\kappa^3 \Big[
 (k_3\cdot h_2\cdot h_1\cdot h_3\cdot k_1)&-(k_3\cdot h_2\cdot h_1\cdot h_3\cdot k_4)
-\frac12(k_1\cdot h_1\cdot k_3)(h_2\cdot h_3)\\
&-(k_2\cdot k_3)(h_1\cdot h_3\cdot h_2)+\text{permutations}(123)\Big]\phi_4
+\text{cyclic}(1234)
~.}}
Due to the properties of the trace and the symmetry of $h$, the term $(h_1\cdot h_3\cdot h_2)$ is invariant under permutations of the positions. Hence the $(k_i\cdot k_j)$ terms 
give a contribution proportional to  $(k_1\cdot k_2+k_2\cdot k_3+k_3\cdot k_1)(h_1\cdot h_3\cdot h_2)$, which vanishes on-shell by momentum conservation. This is consistent with 
the fact that the contribution of \eqref{a41} to $\mathcal{A}_4^{k^2}$ does not contain any terms cubic in $\xi^{(s)}$.

{\it Terms of the form $(k_i\cdot k_j)\phi hb^2 $}

Expanding the one-loop contribution to the $B$-field kinetic term in \eqref{sefffinresc1}, passing to momentum space and keeping terms of the form  $(k_i\cdot k_j)\phi h b^2$ 
we obtain,
\eq{\spl{
-\frac12\gamma\sqrt{G} H^2~\!e^{\sqrt{D-2}~\!\!\phi}\rightarrow
12\gamma \sqrt{D-2}\Big[
(k_1\cdot k_2)(b_1\cdot b_2\cdot h_3)
+\text{cyclic}(123)\Big]\phi _4
+\text{cyclic}(1234)
~.}}
This can be seen to coincide with 
the contribution of \eqref{a41} to $\mathcal{A}_4^{k^2}$, upon expanding the polarization as in \eqref{31}.

{\it Terms of the form $(k_i\cdot k_j)\phi^2 h^2$}

Expanding the quadratic  coupling of the Einstein term to the dilaton around flat space to quadratic order in $h$ and passing to momentum space, we obtain,
\eq{\spl{
\frac12\gamma\sqrt{G} R~\!e^{\sqrt{D-2}~\!\!\phi}\rightarrow
-\gamma (D-2)\Big[
(k_1\cdot k_2)(h_1\cdot h_2)\phi _3\phi _4
+\text{other~pairs}\Big]~.}}
Expanding the polarization as in \eqref{31}, this coincides with the two-$\xi^{(s)}$ contribution of \eqref{a42} to $\mathcal{A}_4^{k^2}$.

{\it Terms of the form $(k_i\cdot k_j)\phi^2 b^2$}

Expanding the one-loop contribution to the $B$-field kinetic term in \eqref{sefffinresc1}, passing to momentum space and keeping terms of the form  $(k_i\cdot k_j)\phi^2 b^2$ 
we obtain,
\eq{\spl{
-\frac12\gamma\sqrt{G} H^2~\!e^{\sqrt{D-2}~\!\!\phi}\rightarrow
-3\gamma (D-2)\Big[
(k_1\cdot k_2)(b_1\cdot b_2)\phi _3\phi _4
+\text{other~pairs}\Big]~.}}
Expanding the polarization as in \eqref{31}, this coincides with the two-$\xi^{(a)}$ contribution of \eqref{a42} to $\mathcal{A}_4^{k^2}$.


\section{Equations of motion and de Sitter solutions}\label{sec:desitter}

The equations of motion following from the effective action \eqref{sefffinresc1} are as follows.

{\it Equation of motion for $G_{\mu\nu}$}
\begin{align}
0=&(1+\gamma e^{\sqrt{D-2}\phi})(R_{rs}-\frac{1}{2}G_{rs}R)-\nabla_r\partial_s(\gamma e^{\sqrt{D-2}\phi})+G_{rs}\nabla^2(\gamma e^{\sqrt{D-2}\phi}) \nonumber \\
&-(1+3\gamma e^{\sqrt{D-2}\phi})e^{-\frac{4}{\sqrt{D-2}}\phi}(H_{r\nu\rho}H_{s}{}^{\nu\rho}  -\frac{1}{6}G_{rs}H^2) \nonumber \\
&-(1+\gamma e^{\sqrt{D-2}\phi})(\partial_r\phi\partial_s\phi-\frac{1}{2}G_{rs}(\partial\phi)^2)-G_{rs} \frac{1}{\alpha'}e^{\frac{2}{\sqrt{D-2}}\phi}(\delta+\alpha'\Lambda e^{\sqrt{D-2}\phi})~,
\end{align}
where we used the variation of the Riemann tensor,
\begin{equation}
\delta R_{mn}=\nabla^s\nabla_{(m}\delta G_{n)s}-\frac{1}{2}\nabla^2\delta G_{mn}-\frac{1}{2}(\nabla_m\nabla_n\delta G_{rs})G^{rs}
~.
\end{equation}
{\it Equation of motion for $\phi$}
\begin{align}
0=&\gamma \sqrt{D-2}e^{\sqrt{D-2}\phi}(\frac{1}{2}R-\frac{1}{2}(\partial \phi)^2) +\nabla^s\left((1+ \gamma e^{\sqrt{D-2}\phi})\partial_s\phi\right) \nonumber \\
&- \gamma \frac{D-6}{2\sqrt{D-2}} H^2e^{\frac{D-6}{\sqrt{D-2}}\phi}+\frac{2}{3\sqrt{D-2}}H^2 e^{-\frac{4}{\sqrt{D-2}}\phi} \nonumber \\
& + \Lambda \frac{D}{\sqrt{D-2}} e^{\frac{D}{\sqrt{D-2}}\phi}
+ \frac{1}{\alpha'}{\frac{2\delta}{\sqrt{D-2}}} e^{\frac{2}{\sqrt{D-2}}\phi}~.
\end{align}
{\it Equation of motion for $b$}
\begin{align}
\nabla^{r}\left((1+3\gamma e^{\sqrt{D-2}\phi})e^{-\frac{4}{\sqrt{D-2}}\phi}H_{rsp}\right)=0
~.
\end{align}
{\it Solutions}

We will look for simple solutions to the equations of motion,  with vanishing fieldstrength for the Kalb-Ramond field, constant dilaton and a maximally symmetric $D$-dimensional space, 
\eq{\phi=\text{constant}~;~~~H=0~;~~~R_{mn}=\lambda G_{mn}~,} 
where $\lambda$ is an arbitrary constant. Then the equation of motion for the $b$-field is automatically satisfied. The equation of motion for the metric reduces to,
\begin{equation}
0=(1+\gamma e^{\sqrt{D-2}\phi}) \frac{D-2}{2}\lambda+\Lambda e^{\frac{D}{\sqrt{D-2}}\phi}+\frac{\delta}{\alpha'}e^{\frac{2}{\sqrt{D-2}}\phi}
~,\label{eqg}
\end{equation}
while the equation of motion for $\phi$ gives,
\begin{align}
\gamma e^{\sqrt{D-2}\phi} \frac{D-2}{2}\lambda=-\Lambda e^{\frac{D}{\sqrt{D-2}}\phi}-\frac{2\delta}{\alpha'D}e^{\frac{2}{\sqrt{D-2}}\phi}~.
\label{eqphi}
\end{align}
By plugging \eqref{eqphi} into \eqref{eqg}, we obtain,
\begin{equation}\label{lam}
\lambda=\tfrac{2(D-26)}{3\alpha'D}~\!e^{\frac{2}{\sqrt{D-2}}\phi}~,
\end{equation}
where we have taken \eqref{dlt} into account.
We can also solve for the dilaton  by substituting this into \eqref{eqphi},
\begin{equation}\label{gs}
e^{\sqrt{D-2}~\!\!\phi}=\frac{2}{\alpha'\Lambda}  
\left( 
c(D-2)+\frac{3D}{(D-26)}
\right)^{-1}~,
\end{equation}
where we have used the definition of $\gamma$, eq.~\eqref{cdef1}. 
Treating $\Lambda$ as a free parameter of the solution, 
we see that $g^2_{\text{str}}= e^{\sqrt{D-2}~\!\!\phi}$ can be made small by taking  $\alpha'\Lambda$ to be sufficiently large. 
The curvature, $\lambda$ is positive for $D>26$, corresponding to de Sitter space. 
Provided $g^2_{\text{str}}$ is small (which, as we mentioned, can be achieved by tuning  $\alpha'\Lambda$ to be sufficiently large), 
the de Sitter space  is weakly curved, $\lambda\alpha'\ll1$, so that the  solution can be trusted.

Even without fine-tuning, i.e.~for $c$, $\Lambda\sim\mathcal{O}(1)$,  we can achieve 
$g^2_{\text{str}}\lesssim 4\%$ for all values of $D$, while $\alpha'\lambda\lesssim10\%$ for $27\leq D\leq32$. 
As a particular numerical example we mention the following solution, 
\eq{\label{nexmpl}
c=\alpha'\Lambda=1~;~~~D=27~;~~~g^2_{\text{str}}=0,019~;~~~\alpha'\lambda=0,018
~.}
It is also interesting to note that in this case  $D\rightarrow\infty$   implies $g^2_{\text{str}}\rightarrow 0$, $\alpha'\lambda\rightarrow\tfrac23$. I.e. in the 
infinite-dimensional limit string perturbation remains valid,  but the supergravity solution becomes strongly curved.

As an alternative example we may use the numerical estimates  \eqref{lest}, \eqref{cest}. 
%
%
These are obtained in the critical dimension, $D=26$, by subtracting the divergent tachyon contribution from the upper strip of the fundamental domain, and performing a ``modified minimal subtraction'' renormalization of the Eisenstein series so as to render $\Lambda$, $c$ finite. We see that $\alpha'\Lambda$ is extremely sensitive to the ratio of gravitational ($l_G$) to string ($l_s$) length. Setting $l_G=l_s$ and extrapolating to $D=27$ gives,
\eq{\label{nexmpl1}
g^2_{\text{str}}=0,00015~;~~~\alpha'\lambda=0,012
~,}
where we used \eqref{lam}, \eqref{gs}. 
Assuming $l_G=10l_S$ instead gives, 
\eq{\label{nexmpl2}
g^2_{\text{str}}=1,57\times10^{-28}~;~~~\alpha'\lambda=0,00014
~.}

\section{Conclusions}\label{sec:conclusions}

We have seen that the one-loop two-derivative effective action remains a well-defined object in the presence of tadpoles and the associated IR divergences, and is rich enough to serve as a starting point for cosmology. 
Moreover, taking a non-supersymmetric string theory such as the bosonic string as a starting point,  sidesteps the question of realizing de Sitter space in the 
context of the effective field theories arising from compactification of the critical ten-dimensional superstrings.

We emphasize that we do not claim to have a realistic model: in particular our treatment simply ignores the tachyonic divergences.  This means that the action \eqref{sefffinresc1}
cannot be taken at face value: rather it should be regarded as a phenomenological model inspired by string theory. As such it is an improvement over the one used in tree-level string cosmology, whose starting point is action \eqref{sefffinresc1} but without the one-loop corrections. 
On the other hand,  the  methods of the present paper can be applied to other nonsupersymmetric models which are tachyon-free and thus do not suffer from the tachyonic pathologies of the bosonic string. Our  results should motivate further investigation in this direction.

In \cite{Tourkine:2013rda} it was argued that the low-energy field theory limit of string theory amplitudes can be systematized by using the language of tropical geometry. 
It would be interesting to apply this formalism to  string theory models with tadpoles and the treatment of the associated IR divergences such as  the ones encountered here.

We have shown that, for $D>26$, the one-loop two-derivative effective action (\ref{sefffinresc1}) admits simple weakly-curved $D$-dimensional de Sitter solutions with constant dilaton at weak string coupling. 
This ensures that, if one were able to solve the issue of the tachyon instability, the solutions would be  in a regime where both the supergravity approximation and the perturbative expansion in the string coupling could be trusted. 
This is in contrast to typical solutions encountered in the context of tree-level string cosmology, where one inevitably reaches a regime where either the string coupling is strong (so that string perturbation cannot be trusted) and/or spacetime is highly curved (so that $\alpha'$ corrections cannot be neglected). Our approach here is also  different from the  supercritical bosonic string models considered in \cite{Hellerman:2006nx}; it is more akin to the approach of \cite{Abel:2015oxa}, in that one argues that the tadpole does not drive the true vacuum 
too far away from flat space, provided the effective (Einstein-frame) cosmological constant is sufficiently small. 

The smallness of the  de Sitter curvature of the solutions presented here is established in the Einstein frame. On the other hand, higher-order derivative corrections to the 
string effective action are controlled by the \emph{string-frame} curvature tensor in units of $\alpha'$. One easily evaluates the  {string-frame} scalar curvature: $\alpha' R=\tfrac{2}{3}(D-26)$, at the de Sitter vacuum of section \ref{sec:desitter}, hence $\alpha' R\geq\tfrac23$ for $D>26$.  
However, the pertinent higher-derivative corrections are not necessarily   controlled by this quantity. 
For example, the  tree-level curvature expansion of the bosonic string effective Lagrangian  is proportional to \cite{Metsaev:1987zx}: 
$e^{-2\varphi}\left\{R+\tfrac{1}{4}\alpha' R_{mnpq}R^{mnpq}+\mathcal{O}(\alpha^{\prime2})\right\}$. The condition of the smallness of the leading-order four-derivative quadratic curvature correction with respect to the Einstein term,   
evaluated at the de Sitter vacuum (in the string frame), can be seen to be equivalent to,
\eq{
1\gg \frac{D-26}{3D(D-1)}
~.}
This is indeed satisfied, as the right-hand side above is smaller than $0,4\%$ for all $D>26$. It would be interesting to check whether  this persists at 
higher loop orders and/or higher orders in derivatives.

Taking the effective action \eqref{sefffinresc1} as the starting point, the requirement for the validity of the de Sitter solutions is a sufficiently large (in string units and in string frame) one-loop vacuum energy. 
This  ensures that the one-loop couplings in \eqref{sefffinresc1} are of the same order as the tree-level ones, while at the same time 
the string coupling is small so that higher-order string loops can be neglected. This is the crucial  feature that allows us 
to avoid having to consider all-order string loop corrections, in contrast to typical studies of higher-loop effects in cosmology \cite{Damour:1994zq,Saharian:1997nu}.  
Moreover the effective cosmological constant (in the Einstein frame) is suppressed by a positive power of the string coupling, ensuring that the typical spacetime curvature is weak. A systematic search for more realistic four-dimensional cosmological solutions is in progress \cite{cosmo} and we hope to report on this in the future.

In the present paper we have treated the one-loop cosmological constant as a freely tunable parameter. This is simply a reflection of our ignorance of its correct renormalized 
value. In section  \ref{sec:desitter} we saw that different  renormalization choices for $\Lambda$ give reasonable values for $\lambda$ and $g^2_{\text{str}}$ at the 
solution.  On the other hand,  in  tachyon-free nonsupersymmetric models, the one-loop cosmological constant has a well-defined finite value, and cannot be freely tuned.  It would be interesting to examine whether such models admit de Sitter solutions in the weakly curved and/or perturbative regime.

As with all non-supersymmetric solutions, 
an important question concerning  the de Sitter solutions presented here is their stability (or absence thereof) with respect to small perturbations. 
An analysis of stability of coupled field perturbations is unfortunately beyond the scope of this paper.  
We hope to return to this  point in the  future.

In deriving the effective action 
\eqref{sefffinresc1}, 
the one-loop corrections were calculated at the critical dimension $D=26$, whereas in order to  include the tree-level cosmological term 
we had to extrapolate off the critical dimension. 
Unfortunately, to our knowledge, there is no systematic way of calculating loop corrections away from the critical dimension. 
Therefore it cannot be excluded that the one-loop part of the action  \eqref{sefffinresc1} should be corrected by terms which vanish in the limit $D\rightarrow26$. 
One might be able to control this ambiguity by considering other nonsupersymmetric models where  the deviation from 
the critical dimension can be taken to be small. It would interesting to examine whether there exist appropriate CFT's which realize this scenario.

\section*{Acknowledgment}

We are grateful to Stefan Hohenegger for comments on the draft, for numerous discussions and ongoing collaboration on related subjects.

\appendix

\section{Various definitions}\label{app:def}

The function $\theta_1$ is given by
\eq{\label{theta}
\theta_1(\nu|\tau)
=-i\sum_{r\in\mathbb{Z}+\frac12}(-1)^{r-\frac12}y^rq^{\frac{r^2}{2}}
~,
}
where $q:=\exp(2i\pi\tau)$, $y:=\exp(2i\pi z)$. It is ``almost'' doubly periodic:
\eq{\label{periodtheta}
\theta_1(z+1|\tau)=-\theta_1(z|\tau)~,~~~
\theta_1(z+\tau|\tau)=-y^{-1}q^{-\frac12}\theta_1(\tau)
~,}
and has the following modular transformations
\eq{\label{modulartheta}
\theta_1(\nu|\tau+1)=\exp(i\pi/4)\theta_1(\nu|\tau)~,~~~
\theta_1(\nu/\tau|-1/\tau)=(-i\tau)^{\frac12}\exp(i\pi\nu^2/\tau)\theta_1(\nu|\tau)
~.}
Dedekind's $\eta$ function is given by,
\eq{\spl{\label{ded}\eta(\tau)&=q^{\frac{1}{24}}\prod_{n=1}^\infty(1-q^n)\\
&=q^{\frac{1}{24}}\sum_{n\in\mathbb{Z}}(-1)^nq^{(3n^2-n)/2}
~,}}
and obeys:
\eq{\label{modulareta}
\eta(\tau+1)=\exp(i\pi/12)\eta(\tau)~,~~~
\eta(-1/\tau)=(-i\tau)^{\frac12}\eta(\tau)
~.}
The  real analytic Eisenstein series is a function of two variables $s$, $\tau$ defined by, 
\eq{\label{e1}E_s(\tau):=
\sum_{m,n\in\mathbb{Z}}^\prime
\frac{\tau_2^s}{|m+\tau n|^{2s}}
~,}
where the prime indicates that the term $(m,n)=(0,0)$ should be omitted from the sum. The series converges for $\text{Re}(s)>1$. It can be analytically continued to a meromorphic function of $s$ on the entire complex plane, 
and has a single pole at $s=1$:
\eq{\label{pole}E_s(\tau)=\frac{\pi}{s-1}+2\pi\left(
\gamma-\ln2-\frac12\ln\tau_2-\ln|\eta(\tau)|^2
\right)+\mathcal{O}(s-1)~.}
Moreover it can be shown that:
\eq{\label{e2}
E_0(\tau)=-1~;~~~\partial_sE_0(\tau)=
-2\left(\ln 2\pi+\frac12\ln\tau_2+\ln|\eta(\tau)|^2
\right)
~.}
\section{Review of general formul\ae{}}\label{sec:amplitude}

The one-loop $N$-point amplitude for the closed bosonic string takes the 
form:
\eq{
\mathcal{A}_N=C g^N\int_F{\text{d}^2\tau}~\!\tau_2^{-13}\left|\eta(\tau)\right|^{-48}
\int_{T^2}\prod_{i=1}^{N-1}{\text{d}^2z_i}\langle V_1(z_1,\bar{z}_1)\cdots V_N(z_N,\bar{z}_N)\rangle
~.\label{ampl}}
For completeness, and in order to fix conventions, let us briefly review the different elements that appear in the formula above, see e.g. \cite{DHoker:1988pdl, Polchinski:1998rq}: 
\begin{itemize}
\item $C$ is an overall normalization that  can be absorbed in the definition of the one-loop vacuum energy, cf.~(\ref{ldef}). $g$ is the 
vertex-operator normalization constant; its relation to the gravitational coupling constant 
$\kappa$ of the 26-dimensional effective action is determined in section \ref{sec:eff}.
\item $\tau:=\tau_1+i\tau_2$ is the modular parameter of the torus $T^2$ and $F$ denotes the fundamental domain,
\eq{-\frac12 \leq \tau_1 \leq \frac12~,~~~|\tau|\geq1~.}
\item The Dedekind function $\eta(\tau)$ is defined in (\ref{ded}). 
In the covariant quantization of the bosonic string, the integration measure of the amplitude  can be understood as coming from functional integration over the reparametrization ghosts. 
\item We integrate over $N-1$ positions $z_i$ of the vertex operators, while we fix $z_N$ by the isometry of $T^2$. The amplitude should be independent of $z_N$.
\item The vertex operators $V_i$ are primary of weight (1,1). In the case of massless particles, they are given by
\eq{\label{masslessv}V_i(z_i,\bar{z}_i)=
 \xi^i_{\mu\nu}\partial X^{\mu}\bar{\partial}X^{\nu}
e^{ik_i\cdot X(z_i,\bar{z}_i)}
~,}
where $X^{\mu}(z_i,\bar{z}_i)$, $\mu=0,1,\dots,25$, are free worldsheet scalars;  $\xi_{\mu\nu}(k_i)$ is the polarization tensor: its symmetric, antisymmetric, trace part describes the graviton $G_{\mu\nu}$, the Kalb-Ramond field (antisymmetric two-form) $b_{\mu\nu}$, the dilaton $\phi$ respectively. 
The condition for $V$ to be primary is equivalent to the on-shell mass condition, $k^2=0$, while the condition for $V$ to have weight (1,1) is the  condition of transversality of the polarization, $k^{\mu}\xi_{\mu\nu}=k^{\nu}\xi_{\mu\nu}=0$.

\item The correlator is given by all possible contractions between vertex operators using the Green's function on the torus,  
\eq{\label{green}G(z)=-\frac{\alpha'}{2}
\log\left|
\frac{2\pi\theta_1(\frac{z}{2\pi}|\tau)}{\theta_1'(0|\tau)}
\right|^2+\alpha'\frac{(\mathrm{Im}z)^2}{4\pi\tau_2}
~.}
This is doubly periodic under $z\rightarrow z+2\pi$, $z\rightarrow z+2\pi\tau$, it has the correct short-distance behavior 
$\lim_{\varepsilon\rightarrow 0}G(\varepsilon)= -\frac{\alpha'}{2}\log\left|\varepsilon\right|^2$, and satisfies, 
\eq{\label{greq}
\frac{1}{\alpha'}\bar{\partial}\partial G(z)=-\pi
\delta^2(z)+\frac{1}{8\pi\tau_2}~.}
The constant ``background charge'' term on the right-hand side above 
is due to the fact that the Laplacian on the torus can only be inverted 
on the space of functions orthogonal to the constant mode;  
we use the conventions of \cite{Polchinski:1998rq} for the various 
normalizations.

\item A regularization scheme must be chosen for the divergent self-contractions 
within each vertex operator. In flat space the self-interactions are subtracted by taking a normal-ordering. More generally for a curved world sheet the regularization can be implemented in a diffeomorphism-invariant way, which however does not manifestly preserve Weyl-invariance. Here we shall adopt the regularization scheme of  \cite{Polchinski:1998rq} according to which for any vertex operator $V$ we define its regularization $[V]_R$ by
\eq{[V]_R:=V+\text{all~selfcontractions~with~}\Delta~;~~~
\Delta(z,z'):=\frac{\alpha'}{2}\ln \text{d}^2(z,z')
~,}
where $\text{d}(z,z')$ is the geodesic distance between the points $z$, $z'$ on 
the worldsheet. For the massless vertex operators (\ref{masslessv}) in particular this gives:
\eq{\spl{\label{regv}
[V_i(z_i,\bar{z}_i)]_R=&
\left( \xi^i_{\mu\nu}\partial X^{\mu}\bar{\partial}X^{\nu}
e^{ik_i\cdot X(z_i,\bar{z}_i)}
-\rm{all~selfcontractions~with~}G\right)\\
&-\frac{\alpha'}{8\pi\tau_2} \xi^i_{\mu}{}^{\mu}\left( 
e^{ik_i\cdot X(z_i,\bar{z}_i)}
-\rm{all~selfcontractions~with~}G\right)
~.}}
In deriving the result above we have taken into account that
\eq{\lim_{z\rightarrow z'}\Delta(z,z')=-\lim_{z\rightarrow z'}G(z-z')~;~~~
\lim_{z\rightarrow z'}\partial\bar{\partial}\Delta(z,z')=-
\partial\bar{\partial}\lim_{z\rightarrow z'}G(z-z')+\frac{\alpha'}{8\pi\tau_2} 
~,}
as follows from (\ref{greq}) and the definition of $\Delta$.

\end{itemize}

In explicit computations of correlators it is useful to make use of the following contractions,
\eq{\spl{
\langle \big[ e^{ik_i\cdot X(z_i,\bar{z}_i)}  \big]_R
 \big[e^{ik_j\cdot X(z_j,\bar{z}_j)}  \big]_R\rangle &=e^{-k_i\cdot k_j G_{ij}}\\
\langle \xi_i\cdot\partial X(z_i,\bar{z}_i)   \big[e^{ik_j\cdot X(z_j,\bar{z}_j)}\big]_R \rangle 
&=i\xi_i\cdot k_j\partial_i G_{ij} \\
\langle  \xi_i\cdot\partial X(z_i,\bar{z}_i)  
 \xi_j\cdot\partial X(z_j,\bar{z}_j)   \rangle &= \xi_i\cdot \xi_j\partial_{i} \partial_{j}G_{ij}\\
\langle  \bar{\xi}_i\cdot\bar{\partial} X(z_i,\bar{z}_i) 
 \xi_j\cdot\partial X(z_j,\bar{z}_j)  \rangle &= \bar{\xi}_i\cdot \xi_j\bar{\partial}_{i} \partial_{j}G_{ij}
~,}}
where $G_{ij}:=G(z_i-z_j)$ and $\partial_i:=\partial/\partial z_i$. 
Note that contrary to the case of the sphere, the $\bar{\partial}X\partial X$ contraction is nonvanishing: this is due to the background charge in (\ref{greq}). Explicitly, using (\ref{green}) we obtain:
\eq{\label{15}\partial_{i} \partial_{j}G_{ij}=
\frac{\alpha'}{8\pi^2}\left(
\frac{\pi}{\tau_2}+\partial_{\nu}\left[
\frac{\theta_1'(\nu|\tau)}{\theta_1(\nu|\tau)}
\right]
\right)
~;~~~
\frac{1}{\alpha'}\partial_{i} \bar{\partial}_{j}G_{ij}=\pi\delta^2(z_i-z_j)-\frac{1}{8\pi\tau_2}
~,
}
where $\nu:=(z_i-z_j)/2\pi$ and $\theta_1'(\nu|\tau):=\partial_{\nu}
\theta_1(\nu|\tau)$. For later use let us also define:
\eq{\label{16}w_{ij}:=\xi_i\cdot \xi_j\partial_{i} \partial_{j}G_{ij}~;~~~
w_{\bar{i}j}:=\bar{\xi}_i\cdot \xi_j\bar{\partial}_{i} \partial_{j}G_{ij}~;~~~
u_{i}:=i\sum_{j\neq i}\xi_i\cdot k_j\partial_i G_{ij}~;~~~
u_{\bar{i}}:=i\sum_{j\neq i}\bar{\xi}_i\cdot k_j\bar{\partial}_i G_{ij}
~,
}
etc, i.e. the expressions with a barred $i$-index are obtained by replacing $(\xi_i,\partial_i)$ with $(\bar{\xi}_i,\bar{\partial}_i)$.

We will use the following representation for the Green's function, 
\eq{\label{grsum}\frac{2\pi}{\alpha'}G(z_1,z_2)=\sum^{\prime}_{
\substack{m,n\in\mathbb{Z}}}
\frac{1}{\lambda_{m,n}}~\!\psi_{m,n}(z_1)\psi_{m,n}^*(z_2)+G_0(\tau)~,}
where $\psi_{m,n}(z)$ is an eigenfunction of eigenvalue $\lambda_{m,n}$ of the Laplacian on the torus; the prime above the sum symbol indicates that the zero eigenvalue $(m,n)=(0,0)$ is excluded; the normalization is chosen so that (\ref{greq}) is obeyed. The zero mode $G_0(\tau)$ does not contribute to the $N$-point amplitude, and will therefore be ignored in the following. Indeed in the correlator of $N$ vertices,  cf.~(\ref{correlt}), the only instance where $G$ does not appear under a derivative is in the term $\sum_{i<j} k_{ij}G_{ij}$, from which $G_0(\tau)$ drops out by virtue of momentum conservation (which implies $\sum_{i<j} k_{ij}=0$).

Explicitly, setting $z:=\sigma_1+\tau\sigma_2$, with $\sigma_i\in[0,1]$, the 
metric of the torus reads:
\eq{\text{d} s^2=\text{d}  z\text{d}  \bar{z}=|\text{d} \sigma_1+\tau\text{d} \sigma_2|^2~,}
with the corresponding Laplacian:
\eq{
\nabla^2=4\partial\bar{\partial}=\frac{1}{\tau_2^2}\left(|\tau|^2\partial_{\sigma_1}^2
-2\tau_1\partial_{\sigma_1}\partial_{\sigma_2}
+\partial_{\sigma_2}^2\right)
~.}
The orthonormal eigenfunctions read:
\eq{
\psi_{m,n}(z)=\frac{1}{\sqrt{\tau_2}}\exp 2\pi i(n\sigma_1+m\sigma_2)
=\frac{1}{\sqrt{\tau_2}}\exp\frac{\pi}{\tau_2}[z(m-n\bar{\tau})-\bar{z}(m-n\tau)]
~,}
with corresponding eigenvalues:
\eq{\lambda_{m,n}=-\frac{4\pi^2}{\tau_2^2}|m-n\tau|^2~.}
Some useful relations which follow immediately from the definition above are:
\eq{\spl{\label{ur}
\psi_{m,n}(-z)&=\psi_{-m,-n}(z)=[\psi_{m,n}(z)]^{*}
=\frac{1}{\tau_2}[\psi_{m,n}(z)]^{-1}\\
\psi_{m,n}(z_{12})&=\sqrt{\tau_2}\psi_{m,n}(z_{1})\psi_{m,n}(-z_{2})\\
\psi_{m,n}(z)\psi_{p,q}(z)&=\frac{1}{\sqrt{\tau_2}}\psi_{m+p,n+q}(z)
~.}}
Inserting the above in (\ref{grsum}) we obtain:
\eq{\spl{\label{g1}\frac{2\pi}{\alpha'}G(z_1,z_2)
&=
\frac{\tau_2}{4\pi^2}\sum_{m,n\in\mathbb{Z}}^{\prime}\frac{1}{|m-n\tau|^2}
\exp\frac{\pi}{\tau_2}[z_{12}(m-n\bar{\tau})-\bar{z}_{12}(m-n\tau)]\\
&=
\frac{\tau_2^{3/2}}{4\pi^2}\sum_{m,n\in\mathbb{Z}}^{\prime}\frac{1}{|m-n\tau|^2}
~\!\psi_{m,n}(z_{12})
~.}}
%
From the above we obtain the following expression for the Green's function at vanishing separation,
\eq{\label{r1}
\frac{2\pi}{\alpha'}\lim_{z_{1}\rightarrow z_{2}} G(z_1,z_2)=\frac{1}{4\pi^2}\lim_{s\rightarrow 1}\sum_{m,n\in\mathbb{Z}}^{\prime}\frac{\tau_2^s}{|m-n\tau|^{2s}}
=\frac{1}{4\pi^2}\lim_{s\rightarrow 1}E_s(\tau)
~,}
where in the last equality we used the definition of the Eisenstein series (\ref{e1}). The limit on the right-hand side above is a pole singularity, 
cf.(\ref{pole}).
From (\ref{g1}) we also obtain:
\eq{\label{uop}\frac{2\pi}{\alpha'}\lim_{z_{1}\rightarrow z_{2}}\partial_1\bar{\partial}_2G(z_1,z_2)=\frac{1}{4\pi^2}\lim_{s\rightarrow 1}\sum_{m,n\in\mathbb{Z}}^{\prime}
\frac{\tau_2^s}{|m-n\tau|^{2s}}\left(\frac{\pi}{\tau_2}\right)^2|m-n\tau|^2
=\frac{1}{4\tau_2}E_0(\tau)=-\frac{1}{4\tau_2}~,}
where we took (\ref{e2}) into account. We see that the regularization above amounts to dropping the delta function in (\ref{15}).

Similarly starting from (\ref{g1}) and using the orthonormality of the Laplacian eigenfunctions on the torus,
\eq{\label{25}
\int_{T^2}\text{d}^2z
~\psi_{m,n}(z)\psi_{-p,-q}(z)
=
\int_{T^2}\text{d}^2z
~\psi_{m,n}(z)\psi^*_{p,q}(z)
=\frac{1}{\sqrt{\tau_2}}\int_{T^2}\text{d}^2z
~\psi_{m-p,n-q}(z)
=\delta_{mp}\delta_{nq}
~,}
we obtain the following useful formula: 
\eq{\spl{\label{utp}
\int_{T^2}\text{d}^2z_1\partial_1\bar{\partial}_2G(z_1,z_2) \bar{\partial}_1{\partial}_2G(z_1,z_2)=\int_{T^2}\text{d}^2z_1\partial_1{\partial}_2&G(z_1,z_2)\bar{\partial}_1\bar{\partial}_2G(z_1,z_2)\\
&=\frac{1}{\tau_2}\left(\frac{\alpha'}{8\pi}\right)^2E_0(\tau)=-\frac{1}{\tau_2}\left(\frac{\alpha'}{8\pi}\right)^2~,}}
where $z_1$ is integrated over the entire area of the torus, 
\eq{T^2:=\left\{\sigma_1+\tau\sigma_2\in\mathbb{C}|(\sigma_1,\sigma_2)
\in[0,1]^2\right\}~,}
and we have taken into account the volume of the torus, 
\eq{\label{tv}\tau_2=\int_{T^2}\text{d}^2z~.} 

\section{Numerical estimates}\label{app:numest}

In this section we will give numerical estimates for the renormalized constants $\Lambda$, $c$.  
The bosonic string one-loop vacuum energy density, $\rho$, in $D=26$ spacetime dimensions is given by \cite{Polchinski:1998rq},
\eq{\spl{\label{a1app}
\rho=-\frac{1}{\kappa^2}\Lambda&=-\frac12(4\pi^2\alpha')^{-13}
\int_F{\text{d}^2\tau}~\!\tau_2^{-14}\left|\eta(\tau)\right|^{-48}\\
&\sim-\frac12(4\pi^2\alpha')^{-13}
\int^{\infty}\!\!\!\!{\text{d}\tau_2}~\!\tau_2^{-14}\left(
 e^{4\pi\tau_2}+24^2+324^2~\!e^{-4\pi\tau_2}+\cdots\right)
~,}}
where in the first equality we took \eqref{sefffinresc} into account and in the last line we used  \eqref{r}. 
This also gives the value of the normalization constant used in \eqref{ampl}, \eqref{ldef}: $C=\tfrac{1}{2}\kappa^2(4\pi^2\alpha')^{-13}$. 
Note that, in our conventions,   
a positive vacuum energy density  corresponds to negative $\Lambda$ and vice-versa.

As explained in \cite{Polchinski:1985zf}, the integrand above can be seen to be identical to the corresponding expression 
for the $D$-dimensional field theory one-loop vacuum energy of bosonic point particles of mass $m_n$,
\eq{\spl{\label{rft}
\rho^{\text{FT}}&=i\sum_{n}\int\frac{\text{d}^Dk}{(2\pi)^D}\int_0^{\infty}\frac{\text{d}l}{2l}
\exp[-(k^2+m_n^2)~\!l/2]\\
&=-\frac12\int_0^{\infty}\frac{\text{d}l}{l}(2\pi l)^{-D/2}\sum_{n}\exp(-m_n^2~\!l/2)
~,}}
provided we set $D=26$ and we identify,
\eq{l=2\pi\alpha'\tau_2~;~~~m_n^2=\tfrac{4}{\alpha'}(n-1)~,~n\in\mathbb{N}
~.}
Of course the $l$-integral in \eqref{rft} is divergent. It can  be regularized by introducing an IR cutoff, $\varepsilon$, so that,  
\eq{
\int_{\varepsilon}^{\infty}\frac{\text{d}l}{2l}
\exp[-(k^2+m_n^2)~\!l/2]=-\frac12\Big[\ln\varepsilon+
\gamma-\ln2+\ln(k^2+m_n^2)+\mathcal{O}(\varepsilon)
\Big]
~,}
where $\gamma=0.57721\dots$ is Euler's constant (note the similarity with \eqref{pole} below). As reviewed in \cite{Polchinski:1998rq}, substituting the ``modified minimal subtraction'' 
renormalization of the integral above, $-\tfrac12\ln(k^2+m_n^2)$, back into \eqref{rft}, is a quick way to arrive at  
the expected field-theory result,
\eq{
\rho^{\text{FT}}=-\frac{i}{2}\sum_{n}\int\frac{\text{d}^Dk}{(2\pi)^D}
\ln(k^2+m_n^2)~.
}
Alternatively one can Wick-rotate the first line of \eqref{rft}, perform the $k^0$ integration which becomes Gaussian, and then integrate $l$ with an IR cutoff.  This  gives,
\eq{
\sum_{n}\int\frac{\text{d}^{D-1}k}{(2\pi)^{D-1}}\left(
-\frac{1}{\sqrt{\varepsilon}}
+\frac12 \sqrt{\vec{k}^2+m_n^2}+\mathcal{O}(\sqrt{\varepsilon})
\right)
~,}
where we have set $k^{\mu}=(k^0,\vec{k})$. 
Applying the same modified minimal subtraction 
renormalization as before then  provides a shortcut to expressing the vacuum energy as a sum of zero-point energies, 
\eq{
\rho^{\text{FT}}=\frac12\sum_{n}\int\frac{\text{d}^{D-1}k}{(2\pi)^{D-1}}  \sqrt{\vec{k}^2+m_n^2}
~.}
Subtracting by hand  the divergent on-shell tachyon contribution from the upper strip (which of course breaks modular invariance),  a numerical integration of the first line of \eqref{a1app} using Mathematica gives,
\eq{\label{lest}
(2\pi l_s)^{26}\rho= -6235,29\dots~;~~~\alpha'\Lambda=\left(\frac{l_G}{l_s} \right)^{24}157,94\dots
~,}
where we have introduced the string and gravitational lengths,\footnote{Recall that in $D$ dimensions the gravitational constant $\kappa$ has engineering dimensions $(\text{length})^{\tfrac{D-2}{2}}$.}
\eq{l_s:=2\pi\sqrt{\alpha'}~;~~~  l_G:=\kappa^{\tfrac{1}{12}}
~,}
respectively. Moreover from \eqref{ldef}, \eqref{r2c}, by applying a modified minimal subtraction renormalization scheme,
\eq{
E_1(\tau)\rightarrow -2\pi \ln \left(\sqrt{\tau_2} |\eta(\tau)|^2 \right)
~,}
cf.~\eqref{pole}, upon numerical integration we obtain,
\eq{\label{cest}c =-0,01319\dots
~,}
where again we have subtracted  the tachyon contribution.

\vskip 1cm

\bibliography{refs}

\begin{thebibliography}{10}

\bibitem{Antoniadis:1988aa}
Ignatios Antoniadis, C.~Bachas, John~R. Ellis, and Dimitri~V. Nanopoulos.
\newblock {Cosmological String Theories and Discrete Inflation}.
\newblock {\em Phys. Lett.}, B211:393--399, 1988.

\bibitem{Antoniadis:1988vi}
Ignatios Antoniadis, C.~Bachas, John~R. Ellis, and Dimitri~V. Nanopoulos.
\newblock {An Expanding Universe in String Theory}.
\newblock {\em Nucl. Phys.}, B328:117--139, 1989.

\bibitem{Antoniadis:1990uu}
Ignatios Antoniadis, C.~Bachas, John~R. Ellis, and Dimitri~V. Nanopoulos.
\newblock {Comments on cosmological string solutions}.
\newblock {\em Phys. Lett.}, B257:278--284, 1991.

\bibitem{Veneziano:1991ek}
G.~Veneziano.
\newblock {Scale factor duality for classical and quantum strings}.
\newblock {\em Phys. Lett.}, B265:287--294, 1991.

\bibitem{Tseytlin:1991xk}
Arkady~A. Tseytlin and C.~Vafa.
\newblock {Elements of string cosmology}.
\newblock {\em Nucl. Phys.}, B372:443--466, 1992.

\bibitem{Tseytlin:1991ss}
Arkady~A. Tseytlin.
\newblock {Dilaton, winding modes and cosmological solutions}.
\newblock {\em Class. Quant. Grav.}, 9:979--1000, 1992.

\bibitem{Tseytlin:1992ye}
Arkady~A. Tseytlin.
\newblock {Cosmological solutions with dilaton and maximally symmetric space in
  string theory}.
\newblock {\em Int. J. Mod. Phys.}, D1:223--245, 1992.

\bibitem{Tseytlin:1992jq}
Arkady~A. Tseytlin.
\newblock {String cosmology and dilaton}.
\newblock In {\em {International Workshop on Theoretical Physics: 6th Session:
  String Quantum Gravity and Physics at the Planck Energy Scale Erice, Italy,
  June 21-28, 1992}}, pages 202--223, 1992.
\newblock [Submitted to: Int. J. Mod. Phys. A(1992)].

\bibitem{Gasperini:1992em}
M.~Gasperini and G.~Veneziano.
\newblock {Pre - big bang in string cosmology}.
\newblock {\em Astropart. Phys.}, 1:317--339, 1993.

\bibitem{Tseytlin:1994cd}
Arkady~A. Tseytlin.
\newblock {On 'rolling moduli' solutions in string cosmology}.
\newblock {\em Phys. Lett.}, B334:315--322, 1994.

\bibitem{Copeland:1994vi}
Edmund~J. Copeland, Amitabha Lahiri, and David Wands.
\newblock {Low-energy effective string cosmology}.
\newblock {\em Phys. Rev.}, D50:4868--4880, 1994.

\bibitem{Easther:1995ba}
Richard Easther, Kei-ichi Maeda, and David Wands.
\newblock {Tree level string cosmology}.
\newblock {\em Phys. Rev.}, D53:4247--4256, 1996.

\bibitem{Tseytlin:1991bu}
Arkady~A. Tseytlin.
\newblock {On the tachyonic terms in the string effective action}.
\newblock {\em Phys. Lett.}, B264:311--318, 1991.

\bibitem{Tseytlin:1988mw}
Arkady~A. Tseytlin.
\newblock {String Theory Effective Action: String Loop Corrections}.
\newblock {\em Int. J. Mod. Phys.}, A3:365--395, 1988.

\bibitem{Gross:1986mw}
David~J. Gross and John~H. Sloan.
\newblock {The Quartic Effective Action for the Heterotic String}.
\newblock {\em Nucl. Phys.}, B291:41--89, 1987.

\bibitem{Green:1999:2}
Michael~B. Green and Pierre Vanhove.
\newblock The low energy expansion of the one-loop type {II} superstring
  amplitude.
\newblock {\em Phys. Rev.}, D61:104011, 2000.

\bibitem{Kawai:1985xq}
H.~Kawai, D.~C. Lewellen, and S.~H.~H. Tye.
\newblock {A Relation Between Tree Amplitudes of Closed and Open Strings}.
\newblock {\em Nucl. Phys.}, B269:1--23, 1986.

\bibitem{Panda:1987md}
Sudhakar Panda.
\newblock {The One Loop Closed Bosonic String Amplitude With External
  Gravitons, Antisymmetric Tensor Fields and Dilatons}.
\newblock {\em Phys. Lett.}, B193:225--232, 1987.

\bibitem{Abe:1989yb}
Mitsuko Abe.
\newblock {The Modular Invariant Regularization Method and One-Loop Corrected
  Effective Action in the Closed Bosonic String in D =26}.
\newblock {\em Prog. Theor. Phys.}, 82:804, 1989.

\bibitem{Minahan:1989cb}
Joseph~A. Minahan.
\newblock {Calculation of the One Loop Graviton Mass Shift in Bosonic String
  Theory}.
\newblock {\em Nucl. Phys.}, B333:525--535, 1990.

\bibitem{Abel:2015oxa}
Steven Abel, Keith~R. Dienes, and Eirini Mavroudi.
\newblock {Towards a nonsupersymmetric string phenomenology}.
\newblock {\em Phys. Rev.}, D91(12):126014, 2015.

\bibitem{Polchinski:1998rq}
J.~Polchinski.
\newblock {\em {String theory. Vol. 1: An introduction to the bosonic string}}.
\newblock Cambridge University Press, 2007.

\bibitem{Berg:2016wux}
Marcus Berg, Igor Buchberger, and Oliver Schlotterer.
\newblock {From maximal to minimal supersymmetry in string loop amplitudes}.
\newblock {\em JHEP}, 04:163, 2017.

\bibitem{Minahan:1987ha}
Joseph~A. Minahan.
\newblock {One Loop Amplitudes on Orbifolds and the Renormalization of Coupling
  Constants}.
\newblock {\em Nucl. Phys.}, B298:36--74, 1988.

\bibitem{DHoker:2015gmr}
Eric D'Hoker, Michael~B. Green, and Pierre Vanhove.
\newblock {On the modular structure of the genus-one Type II superstring low
  energy expansion}.
\newblock {\em JHEP}, 08:041, 2015.

\bibitem{Tourkine:2013rda}
Piotr Tourkine.
\newblock {Tropical Amplitudes}.
\newblock {\em Annales Henri Poincare}, 18(6):2199--2249, 2017.

\bibitem{Hellerman:2006nx}
Simeon Hellerman and Ian Swanson.
\newblock {Cosmological solutions of supercritical string theory}.
\newblock {\em Phys. Rev.}, D77:126011, 2008.

\bibitem{Damour:1994zq}
T.~Damour and Alexander~M. Polyakov.
\newblock {The String dilaton and a least coupling principle}.
\newblock {\em Nucl. Phys.}, B423:532--558, 1994.

\bibitem{Saharian:1997nu}
Aram~A. Saharian.
\newblock {Higher loop string cosmology with moduli and antisymmetric tensor
  field}.
\newblock {\em Class. Quant. Grav.}, 15:1951--1970, 1998.

\bibitem{cosmo}
S~Hohenegger and D.~Tsimpis.
\newblock In progress.

\bibitem{DHoker:1988pdl}
Eric D'Hoker and D.~H. Phong.
\newblock {The Geometry of String Perturbation Theory}.
\newblock {\em Rev. Mod. Phys.}, 60:917, 1988.

\bibitem{Polchinski:1985zf}
Joseph Polchinski.
\newblock {Evaluation of the One Loop String Path Integral}.
\newblock {\em Commun. Math. Phys.}, 104:37, 1986.

\end{thebibliography}
\bibliographystyle{unsrt}
\end{document}